%% file: BLADE.tex
\documentclass[sigconf,natbib=true]{acmart}

\usepackage{enumitem}
\setlist[itemize]{leftmargin=*}
\usepackage{algorithm}
\usepackage{algorithmic}

\usepackage[table]{xcolor}
\usepackage{subcaption}

\usepackage{colortbl}
\usepackage{multirow}
\usepackage{graphicx}
\usepackage[normalem]{ulem}
\useunder{\uline}{\ul}{}

\definecolor{gblue}{RGB}{66,133,244}

\usepackage{arydshln}
\makeatletter
\def\adl@drawiv#1#2#3{%
        \hskip.5\tabcolsep
        \xleaders#3{#2.5\@tempdimb #1{1}#2.5\@tempdimb}%
                #2\z@ plus1fil minus1fil\relax
        \hskip.5\tabcolsep}
\newcommand{\cdashlinelr}[1]{%
  \noalign{\vskip\aboverulesep
           \global\let\@dashdrawstore\adl@draw
           \global\let\adl@draw\adl@drawiv}
  \cdashline{#1}
  \noalign{\global\let\adl@draw\@dashdrawstore
           \vskip\belowrulesep}}
\makeatother

\usepackage[normalem]{ulem} 

\usepackage{colortbl} 
\definecolor{mygray}{rgb}{0.9, 0.9, 0.9}
\definecolor{myred}{rgb}{0.68627451, 0.14117647, 0.09803922}
\newcommand{\myeq}[1]{{Eq.~(\ref{eq:#1})}}
\newcommand{\mysec}[1]{{Section~\ref{sec:#1}}}
\newcommand{\mytab}[1]{{Table~\ref{tab:#1}}}
\newcommand{\myfig}[1]{{Fig.~\ref{fig:#1}}}

\AtBeginDocument{%
  }

\copyrightyear{2026}
\acmYear{2026}
\setcopyright{cc}
\setcctype{by}
\acmConference[SIGIR '26]{Proceedings of the 49th International ACM SIGIR Conference on Research and Development in Information Retrieval}{July 20--24, 2026}{Melbourne, VIC, Australia}
\acmBooktitle{Proceedings of the 49th International ACM SIGIR Conference on Research and Development in Information Retrieval (SIGIR '26), July 20--24, 2026, Melbourne, VIC, Australia}
\acmDOI{10.1145/3805712.3809535}
\acmISBN{979-8-4007-2599-9/2026/07}

\begin{document}

\title{Beyond Static Best-of-N: Bayesian List-wise Alignment for LLM-based Recommendation}


\author{Ruijun Chen}
\orcid{0009-0009-0186-9561}
\email{rjchen20@mail.ustc.edu.cn}
\affiliation{%
  \institution{University of Science and Technology of China}
  \city{Hefei}
  \country{China}
}

\author{Chongming Gao}
\authornotemark[1]
\email{chongming.gao@gmail.com}
\affiliation{%
  \institution{University of Science and Technology of China}
  \city{Hefei}
  \country{China}
}
\orcid{0000-0002-5187-9196}

\author{Jiawei Chen}
\email{sleepyhunt@zju.edu.cn}
\affiliation{%
  \institution{Zhejiang University}
  \city{Hangzhou}
  \country{China}
}
\orcid{0000-0002-4752-2629}

\author{Weiqin Yang}
\email{futurelover10032@gmail.com}
\affiliation{%
  \institution{Zhejiang University}
  \city{Hangzhou}
  \country{China}
}
\orcid{0000-0002-5750-5515}

\author{Xiangnan He}
\authornote{Corresponding Author.}
\email{xiangnanhe@gmail.com}
\orcid{0000-0001-8472-7992}
\affiliation{%
\institution{University of Science and Technology of China}
\city{Hefei}
\country{China}
}

\renewcommand{\shortauthors}{Ruijun Chen et al.}

\begin{abstract}
Large Language Models have revolutionized recommender systems (LLM4Rec) by leveraging their generative capabilities to model complex user preferences. However, existing LLM4Rec methods primarily rely on token-level objectives, making it difficult to optimize list-level and non-differentiable metrics (e.g., NDCG, fairness) that define actual recommendation quality. While Best-of-N (BoN) directly optimizes these metrics during inference, its high computational cost hinders real-world deployment. To address this, BoN Alignment aims to distill the search capability into the model itself, yet current approaches suffer from two critical limitations: (1) \textbf{Indiscriminate Supervision}, where the static reference fails to distinguish the relative quality of candidates exceeding its empirical range, leading to a loss of ranking guidance; and (2) \textbf{Gradient Decay}, where the effective supervision signal rapidly diminishes as the evolving policy improves, resulting in inefficient optimization.

To overcome these challenges, we propose \textbf{BLADE} (\textbf{B}ayesian \textbf{L}ist-wise \textbf{A}lignment via \textbf{D}ynamic \textbf{E}stimation). Unlike static approaches, BLADE introduces a Bayesian framework that continuously updates the target distribution by fusing historical priors with dynamic evidence from the model's current rollouts. This mechanism constructs a self-evolving target that adapts to the model's growing capabilities, ensuring the training signal remains informative throughout the learning process. Extensive experiments on three real-world datasets demonstrate that BLADE significantly outperforms state-of-the-art baselines. Crucially, it breaks the static performance upper bound, achieving sustained gains in both ranking accuracy (Recall, NDCG) and complex list-wise metrics (Fairness, Diversity). The code is available via \url{https://github.com/RegionCh/BLADE}.
\end{abstract}




\maketitle

\input{sections/1.Introduction}
\input{sections/2.Preliminary}

\input{sections/3.Method}
\input{sections/4.Experiments}
\input{sections/5.Related}


\section{Conclusion}

In this work, we addressed the critical limitation of standard LLM4Rec methods, which rely on token-level objectives and consequently struggle to optimize complex, non-differentiable list-wise metrics. While BoN alignment offers a potential solution, we identified two fundamental bottlenecks in its static implementation: \textbf{Indiscriminate Supervision}, where the static reference fails to distinguish superior candidates and \textbf{Gradient Decay} due to diminishing supervision signals. To overcome these challenges, we proposed \textbf{BLADE}, a \textbf{B}ayesian \textbf{L}ist-wise \textbf{A}lignment framework via \textbf{D}ynamic \textbf{E}stimation. By reformulating the alignment target as a self-evolving posterior that fuses historical priors with real-time dynamic evidence, BLADE ensures that the optimization process remains informative and effective throughout the training trajectory. Extensive experiments demonstrate that BLADE not only significantly outperforms state-of-the-art baselines in ranking accuracy but also effectively breaks the performance upper bound imposed by static references. Furthermore, our analysis confirms that BLADE serves as a robust, metric-agnostic framework capable of navigating the Pareto frontier between accuracy and complex list-wise objectives such as fairness and diversity.


\section*{Acknowledgements}
This work is supported by the National Natural Science Foundation of China (62402470,U24B20180,62525211), the Fundamental Research Funds for the Central Universities of China (WK2100000053), Anhui Provincial Natural Science Foundation (2408085QF189). This research is supported by the advanced computing resources provided by the Supercomputing Center of the USTC.

\bibliographystyle{ACM-Reference-Format}
\balance
\small
\bibliography{BLADE}

\end{document}

%% file: sections/1.Introduction.tex
\section{Introduction}

The integration of Large Language Models (LLMs) into recommender systems, known as LLM4Rec~\cite{lin2025can,surveyLLM4rec}, leverages their semantic reasoning and generalization capabilities to enhance personalization~\cite{LLaRA,dai2025onepiece}. Typically, these methods serialize user historical interactions into natural language prompts, instructing the model to generate the titles of target items sequentially~\cite{geng2022recommendation}. Since pre-trained LLMs often lack the domain-specific knowledge required to generate valid recommendations~\cite{bao2023tallrec}, current research primarily relies on Supervised Fine-Tuning (SFT)~\cite{bao2023bi} and alignment techniques like Direct Preference Optimization (DPO)~\cite{rafailov2024direct,chen2024softmax}. However, these approaches are predominantly grounded in token-level likelihoods or pair-wise preferences, creating a critical misalignment with list-wise objectives. In real-world scenarios, user satisfaction fundamentally relies on holistic list-level metrics such as NDCG and fairness, which capture global item interactions~\cite{zhao2017deep}. Consequently, optimization methods restricted to local token objectives fail to directly maximize these intricate, non-differentiable rewards.

\begin{figure}[t]
    \centering
    
    \begin{subfigure}{0.62\linewidth}
        \centering
        \includegraphics[width=\linewidth]{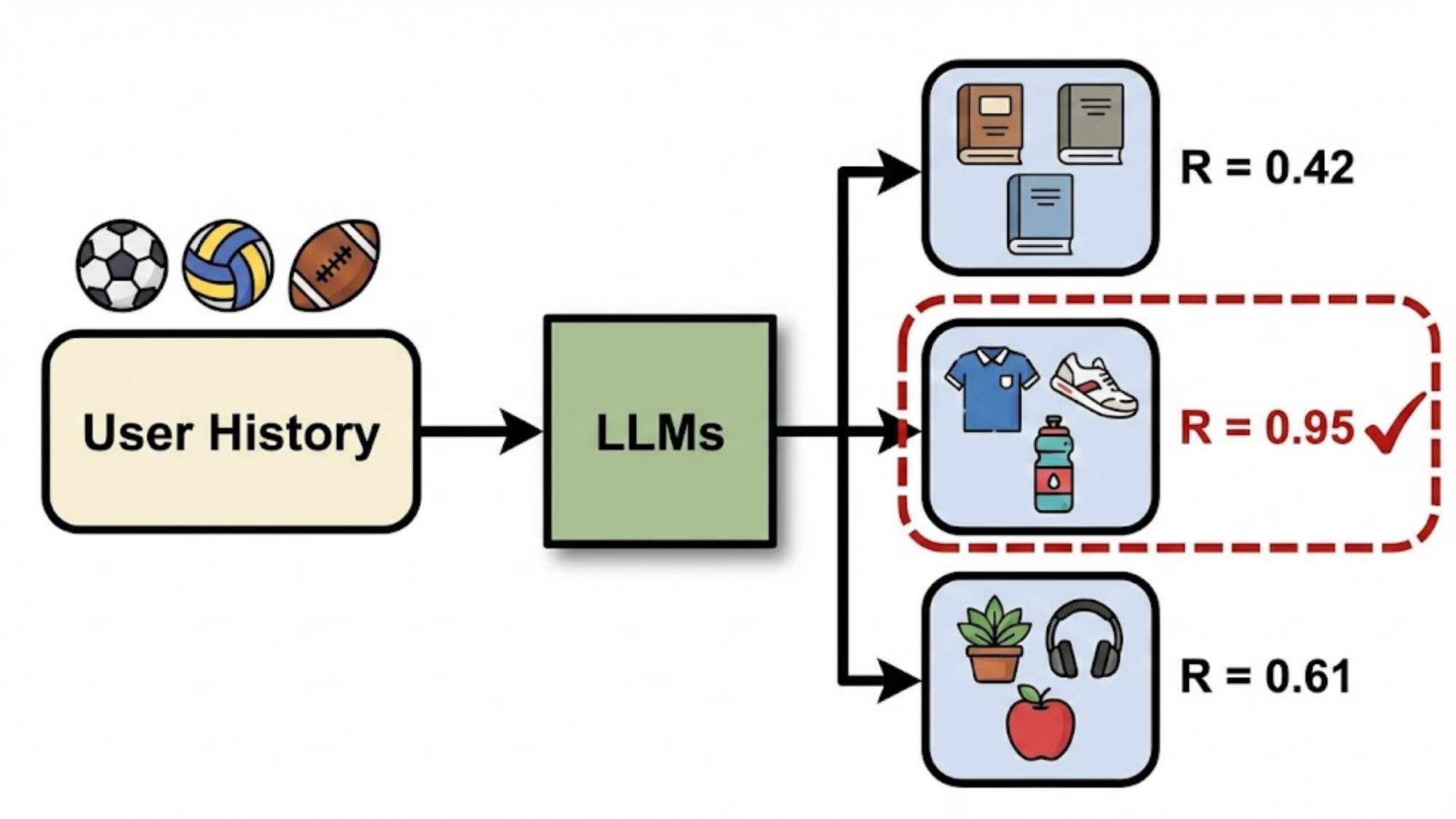} 
        \caption{Inference-time Best-of-N}
        \label{fig:bon_method} 
    \end{subfigure}
    \hfill 
    \begin{subfigure}{0.35\linewidth}
        \centering
        \includegraphics[width=\linewidth]{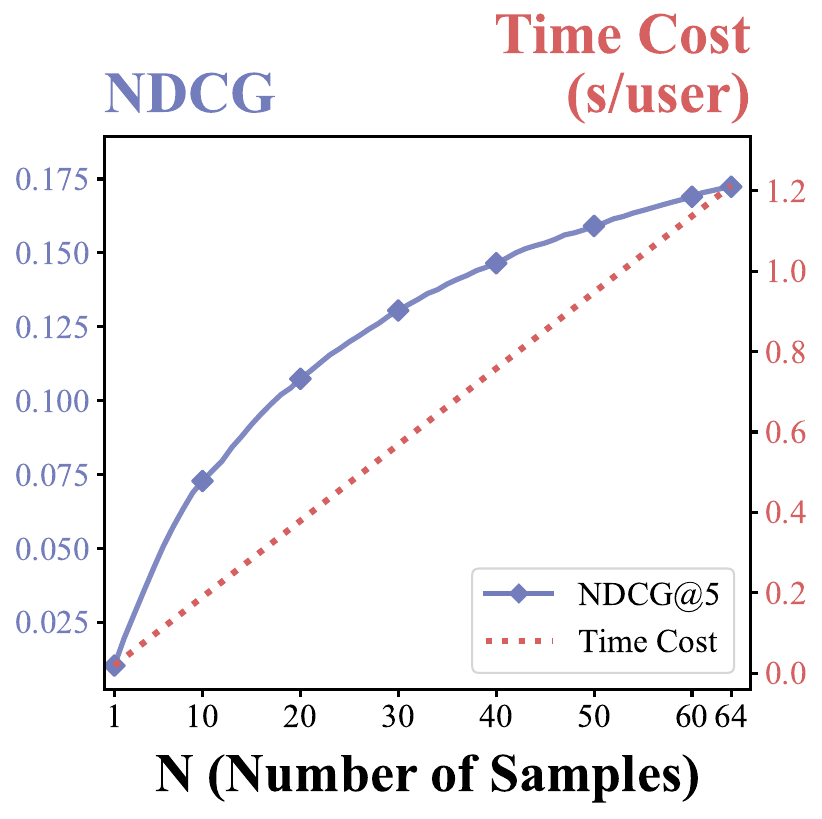}
        \caption{NDCG vs. Latency}
        \label{fig:ndcg_latency}
    \end{subfigure}
    
    \caption{
        The efficiency bottleneck of Best-of-N in LLM4Rec.
        \subref{fig:bon_method} BoN selects the optimal candidate from $N$ samples to maximize list-wise rewards.
        \subref{fig:ndcg_latency} However, increasing $N$ linearly escalates inference latency, making real-time deployment impractical.
    }
    \label{fig:BoN_bottleneck}
\end{figure}

Given the challenges of optimizing these non-differentiable rewards directly via gradients, inference-time search has emerged as a robust alternative~\cite{welleck2024decoding,wei2022chain}. Specifically, Best-of-N (BoN)~\cite{stiennon2020learning} effectively bypasses the differentiability constraint by sampling $N$ candidates and selecting the optimal one based on the list-wise reward (as shown in \myfig{bon_method}). This strategy has demonstrated substantial performance gains in aligning LLMs~\cite{snell2024scaling,wu2024inference}. However, this effectiveness comes at a prohibitive cost. As illustrated in \myfig{ndcg_latency}, the inference latency increases linearly with the sample size $N$, rendering standard BoN impractical for online serving where low latency is mandated. This dilemma motivates the adoption of \textit{BoN Alignment}~\cite{amini2024variational,sessa2024bond}, which aims to internalize the search-and-selection capability into the model’s weights, enabling the direct generation of high-quality responses in a single pass.

Despite the promise of BoN alignment, its application to list-level recommendation faces two critical limitations stemming from the reliance on a static reference distribution. First, it is hindered by \textbf{Indiscriminate Supervision}. Since the static reference relies on a fixed support, its reward quantile estimation saturates for candidates outperforming the reference. Consequently, the alignment objective fails to distinguish between these superior candidates, assigning them identical target probabilities and thereby stripping the policy of critical ranking guidance in high-reward regions. Second, we identify the issue of \textbf{Gradient Decay}. As the policy evolves and rapidly outpaces the static reference, the proportion of reference samples offering informative signals drastically shrinks. This leads to a vanishing supervision signal, causing the optimization dynamics to stall prematurely.

To overcome these challenges, we propose \textbf{BLADE} (\textbf{B}ayesian \textbf{L}ist-wise \textbf{A}lignment via \textbf{D}ynamic \textbf{E}stimation). Rather than anchoring alignment to a fixed distribution, BLADE introduces a Bayesian framework to dynamically integrate static reference priors with \textit{dynamic evidence} derived from the model's current rollouts. This enables the online estimation of the evolving BoN distribution, reformulating the learning objective into a \textit{self-evolving target}. Such dynamic alignment ensures that the training signal remains informative as the model's capability expands, enabling continuous optimization that transcends the limitations of static reference models. To validate our approach, we conduct extensive experiments on three real-world datasets, demonstrating that BLADE achieves sustained improvements across both standard ranking accuracy and complex list-wise metrics.

To summarize, our main contributions are as follows:

\begin{itemize}
    \item We systematically investigate the potential of BoN strategies in LLM4Rec and identify two critical bottlenecks in static BoN alignment: \textbf{Indiscriminate Supervision} stemming from the reference's inability to distinguish superior candidates, and \textbf{Gradient Decay} caused by diminishing supervision signals.
    
    \item We propose \textbf{BLADE}, a Bayesian framework that dynamically fuses the static reference with dynamic evidence, constructing a \textit{self-evolving target} to ensure sustained, informative gradients throughout training.
    
    \item Experiments on three datasets demonstrate that BLADE breaks the static performance upper bound, significantly outperforming baselines in both ranking utility and complex list-wise objectives such as fairness and diversity.
\end{itemize}

%% file: sections/2.Preliminary.tex
\section{Preliminaries}
\label{sec:preliminary}

In this section, we introduce the theoretical foundation of our work. We first formulate the recommendation task as a conditional list generation problem. Subsequently, we introduce GRPO, which serves as the optimization backbone of our framework. We conclude by  reviewing the Best-of-N strategy, detailing its application in both inference and policy alignment contexts.

\subsection{Task Formulation}
We formulate the recommendation task as a conditional sequence generation problem. Let $\mathcal{U}$ and $\mathcal{I}$ denote the sets of users and items, respectively. For each user $u \in \mathcal{U}$, we construct a natural language prompt $x$ based on their interaction history. The LLM functions as a stochastic policy $\pi_\theta$, which takes $x$ as input and autoregressively generates a response $y = (y_1, y_2, \dots, y_T)$ consisting of $T$ tokens. This response $y$ represents an ordered list of $K$ recommended items. The probability of generating the complete response is factorized as:
\begin{equation}
    \pi_\theta(y | x) = \prod_{t=1}^{T} \pi_\theta(y_t | x, y_{<t}).
    \label{eq:policy}
\end{equation}
The optimization goal is to maximize a non-differentiable list-wise reward $R(y; u)$, such as NDCG or Diversity, which evaluates the quality of the item sequence embedded within the generated response $y$.

\subsection{GRPO}
\label{sec:grpo_prelim}

Standard Reinforcement Learning methods like PPO typically rely on a critic to estimate expected returns, which doubles the memory footprint and introduces training instability due to value approximation errors. To circumvent these issues, we adopt Group Relative Policy Optimization (GRPO)~\cite{Shao2024DeepSeekMathPT}.

Formally, in the recommendation scenario, for a given user context $x$, the policy $\pi_{\theta}$ samples $G$ independent candidate lists $\{y_1, y_2, \dots, y_G\}$. Rather than relying on an external reward model, GRPO leverages the relative quality of generations within the group. For any specific list-wise reward function $R(y)$ (e.g., NDCG or a customized metric), the advantage $A_i$ for the $i$-th candidate is computed by normalizing its score against the group statistics:
\begin{equation}
    A_i = \frac{R(y_i) - \mu_G}{\sigma_G},
    \label{eq:grpo_advantage}
\end{equation}
where $\mu_G$ and $\sigma_G$ denote the mean and standard deviation of the rewards $\{R(y_j)\}_{j=1}^G$ within the current group.

The policy is then updated by maximizing the surrogate objective, which incorporates a clipped probability ratio to prevent destructive updates. The objective function is defined as:
\begin{equation}
\begin{aligned}
    \mathcal{J}(\theta) = \mathbb{E}_{\pi_{\theta_{\text{old}}}} \Bigg[ \frac{1}{G} \sum_{i=1}^G \frac{1}{T_i} \sum_{t=1}^{T_i} \bigg( & \min \left( \rho_{i,t} A_i, \text{clip}(\rho_{i,t}, 1-\epsilon, 1+\epsilon) A_i \right) \\
    & - \beta \mathbb{D}_{\text{KL}} \left( \pi_\theta(\cdot|h_{<t}) \parallel \pi_{\text{ref}}(\cdot|h_{<t}) \right) \bigg) \Bigg]
\end{aligned}
\label{eq:grpo_objective}
\end{equation}
where $T_i$ is the length of response $y_i$, and $\rho_{i,t} = \frac{\pi_{\theta}(y_{i,t}|x, y_{i,<t})}{\pi_{\theta_{\text{old}}}(y_{i,t}|x, y_{i,<t})}$ is the token-level importance sampling ratio.

In the context of our work, GRPO serves as an efficient backbone optimization algorithm. By leveraging group-based relative feedback, it enables us to directly optimize non-differentiable list-wise metrics without the computational overhead of a separate critic network. This stability and memory efficiency make it an ideal foundation for implementing our proposed dynamic alignment framework.

\subsection{Best-of-N Framework}
\label{sec:bon_framework}

\subsubsection{Best-of-N Inference}
To directly optimize non-differentiable list-wise metrics, \textbf{BoN sampling}~\cite{stiennon2020learning} serves as a powerful strategy. Specifically, in the context of recommendation, given a user context $x$, the policy $\pi_{\theta}$ generates $N$ independent candidate recommendation lists $\{y_1, \dots, y_N\}$. The system then selects the optimal list $y^*$ by maximizing the list-wise reward $R(y)$:
\begin{equation}
    y^* = \operatorname*{argmax}_{y \in \{y_1, \dots, y_N\}} R(y).
\end{equation}
While this approach effectively bypasses gradient limitations, it incurs a substantial computational cost. As demonstrated in Table~\ref{tab:main_results}, although increasing $N$ significantly boosts performance, it results in a linear increase in inference latency, rendering it impractical for real-time recommendation.

\subsubsection{Best-of-N Alignment}
To mitigate the computational bottleneck of sampling, BoN Alignment aims to internalize the search capability into the policy's parameters.
Following the variational perspective of vBoN~\cite{amini2024variational}, the exact probability mass that a sample $y$ is selected by BoN sampling is given by the expansion of order statistics:
\begin{equation}
\label{bon_distribution}
    \pi_{\text{BoN}}(y) = \sum_{k=1}^{N} \binom{N}{k} 
    \left(F_{\text{ref}}(R(y))\right)^{N-k} \pi_{\text{ref}}(y)^k,
\end{equation}
where $F_{\text{ref}}(r) = P_{y' \sim \pi_{\text{ref}}}(R(y') \le r)$ denotes the Cumulative Distribution Function (CDF) of the reward under the reference policy.
Here, the summation accounts for the fact that the same recommendation list $y$ may appear multiple times among the $N$ i.i.d. samples.
The term indexed by $k$ corresponds to the event that $y$ is sampled $k$ times, while the remaining $N-k$ samples have rewards no larger than $R(y)$.

Directly optimizing the exact BoN distribution is computationally inconvenient, since the reverse-KL objective involves the logarithm of the above order-statistic summation.
This variational formulation therefore introduces a tractable lower-bound surrogate, yielding a quantile-based reward together with KL regularization:
\begin{equation}
\label{eq:vbon_surrogate}
    \mathcal{J}_{\text{BoN}}(\theta)
    =
    \mathbb{E}_{y \sim \pi_\theta}
    \left[
        (N-1)\log F_{\text{ref}}(R(y))
        - \log \frac{\pi_\theta(y)}{\pi_{\text{ref}}(y)}
    \right].
\end{equation}

Consequently, aligning the policy $\pi_\theta$ with this target is equivalent to minimizing the KL divergence, which translates to maximizing the following objective:
\begin{equation}
\label{eq:implicit_reward}
\begin{aligned}
    \mathcal{J}(\theta) &= -\mathbb{D}_{\text{KL}}(\pi_{\theta} || \pi_{\text{BoN}}) \\
    &= \mathbb{E}_{y \sim \pi_{\theta}} \left[ \log \frac{\pi_{\text{BoN}}(y)}{\pi_{\theta}(y)} \right] \\
    &\simeq \mathbb{E}_{y \sim \pi_{\theta}} \bigg[ \underbrace{(N-1) \log F_{\text{ref}}(R(y))}_{\text{Quantile Reward}} - \underbrace{\log \frac{\pi_{\theta}(y)}{\pi_{\text{ref}}(y)}}_{\text{KL Regularization}} \bigg].
\end{aligned}
\end{equation}
Here, the objective implies an RL problem where the model is incentivized to generate responses with high reward quantiles relative to the reference, while remaining anchored to the reference via KL regularization.

In standard implementations, the CDF term $F_{\text{ref}}(\cdot)$ is estimated empirically using a fixed set of samples from a \textbf{static reference model}. While computationally convenient, relying on a fixed anchor introduces fundamental theoretical limitations regarding \textbf{ranking discrimination} and gradient dynamics. We formally analyze these bottlenecks in \mysec{method} to motivate the design of our proposed BLADE framework.

\begin{table}[h]
\centering
\caption{Comparison of recommendation performance (NDCG@5) and inference efficiency between BLADE, BoN Alignment, and Inference-time BoN. }
\label{tab:main_results}
\begin{tabular}{lccccc}
\toprule
\textbf{Method}  & \textbf{NDCG@5} & \textbf{Time (s)} & \textbf{Inference Cost} \\
\midrule
Base  & 0.0283 & \textbf{44.56} & $1\times$ \\
BoN ($N=2$) &  0.0380 & 89.12 & $2\times$ \\
BoN ($N=4$) &  \textbf{0.0470} & 178.24 & $4\times$ \\
BoN Alignment  & 0.0333 & \textbf{44.56} & $1\times$ \\
\midrule
\textbf{BLADE} &  \underline{0.0410} & \textbf{44.56} & \textbf{$1\times$} \\
\bottomrule
\end{tabular}
\end{table}

%% file: sections/3.Method.tex
\section{BLADE}
\label{sec:method}

In this section, we present \textbf{BLADE}. We begin by analyzing the fundamental limitations of static BoN alignment. Subsequently, we outline the overall framework, detail the Bayesian formulation for dynamic target estimations, and finally describe the optimization process using GRPO.

\begin{figure}
    \centering
    \includegraphics[width=\linewidth]{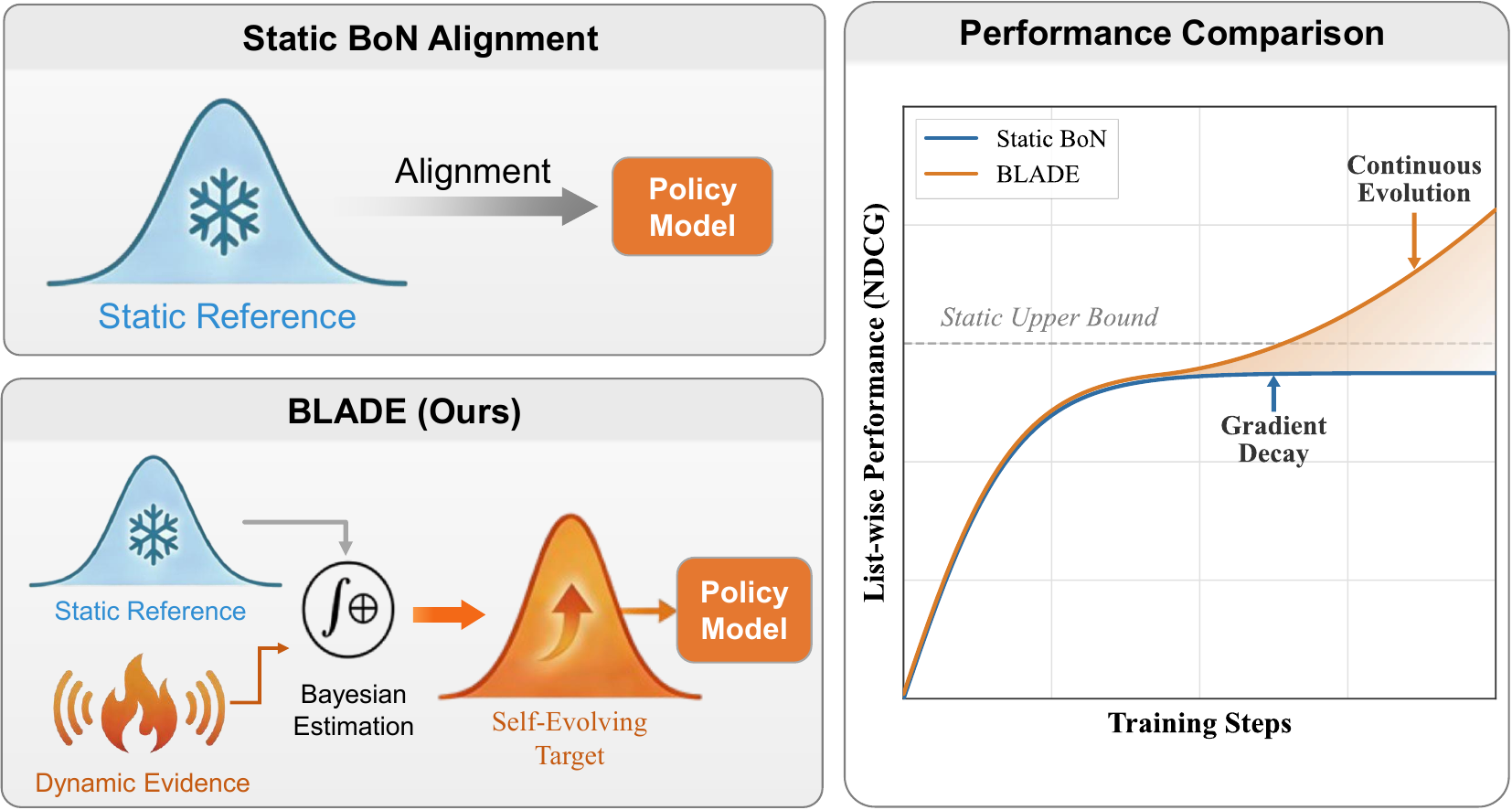}
    \caption{Static BoN Alignment vs. BLADE. (Left) Unlike static methods anchored to a frozen reference, BLADE constructs a Self-Evolving Target by fusing static priors with dynamic evidence, thereby resolving Indiscriminate Supervision in high-reward regions. (Right) This mechanism mitigates Gradient Decay, enabling the model to break the Static Upper Bound for continuous evolution.}
    \label{fig:comparison_bon_blade}
\end{figure}

\subsection{Limitations of Static Reference}
\label{sec:limitations}
Despite the tractability of the approximation in Eq.~(\ref{eq:implicit_reward}), existing methods typically estimate the target distribution using a \textbf{static} reference model. We formally analyze how this static nature imposes two fundamental limitations.

\begin{figure*}[t]
    \centering
    \begin{minipage}[b]{0.635\textwidth}
        \centering
        \includegraphics[width=\linewidth]{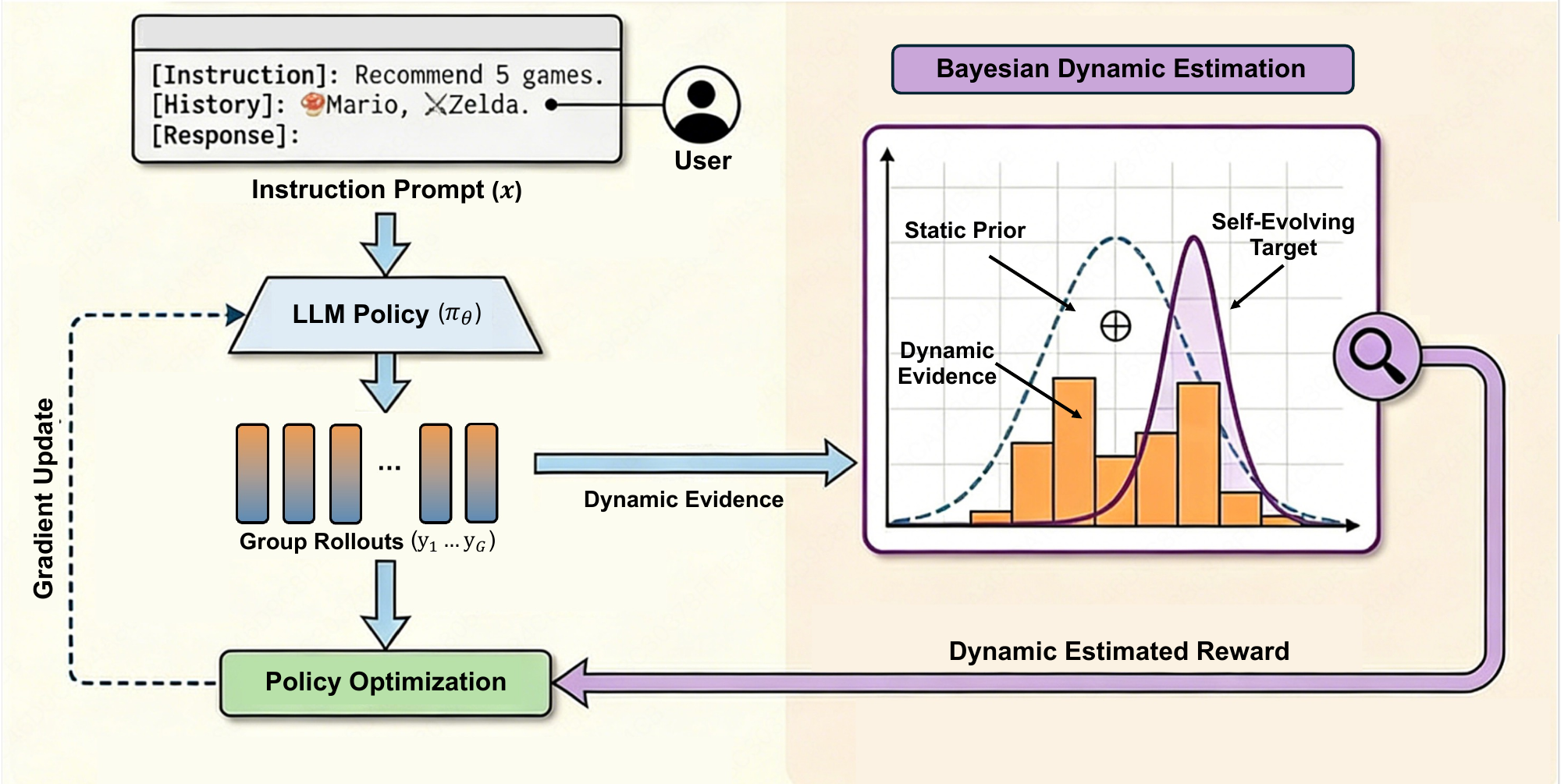}
    \end{minipage}%
    \hfill
    \begin{minipage}[b]{0.34\textwidth}
        \centering
        \small 
        \begin{algorithm}[H]
            \caption{BLADE Training Algorithm}
            \label{alg:blade_final}
            \begin{algorithmic}[1]
                \REQUIRE Policy $\pi_{\theta}$, Reference $D_{ref}$, $\tau, N$
                \STATE Pre-compute static statistics $\{N_{ref}^{<r}\}$
                \FOR{each batch $X \sim \mathcal{D}$}
                    \STATE Generate candidates $\{y_{i}\}_{i=1}^G \sim \pi_{\theta}(\cdot|X)$
                    
                    \FOR{each candidate $y_i$}
                        \STATE \textbf{1. Dynamic Evidence:} 
                        \STATE \quad $N_{batch}^{<R(y_i)} \leftarrow \sum_{j=1}^G \mathbb{I}(R(y_j) < R(y_i))$
                        \STATE \textbf{2. Estimate Target} (Bayesian Fusion):
                        \STATE \quad $\hat{F}_i \leftarrow \frac{N_{ref}^{<R(y_i)} + \tau \cdot N_{batch}^{<R(y_i)}}{ |D_{ref}| + \tau \cdot G }$
                        \STATE \textbf{3. Proxy Reward:} $r_i \leftarrow (N-1) \log \hat{F}_i$
                    \ENDFOR
                    
                    \STATE \textbf{4. Update:} Optimization step via GRPO
                \ENDFOR
            \end{algorithmic}
        \end{algorithm}
    \end{minipage}
    \vspace{-0.2cm}
    \caption{The overall training framework of BLADE. 
        (Left) The pipeline illustrates how BLADE fuses the Static Prior with Dynamic Evidence to construct a self-evolving target distribution.
        (Right) The detailed training algorithm. For each batch, BLADE computes the Bayesian dynamic target $\hat{F}_i$ via \myeq{blade_formula} and converts it into a proxy reward for list-wise policy optimization.}
    \label{fig:framework}
\end{figure*}

\paragraph{1. Indiscriminate Supervision}
Recall that the BoN target distribution is derived as a re-weighting of the reference. In standard implementations, the CDF $F_{\text{ref}}(\cdot)$ is empirically estimated using a fixed set of samples $\mathcal{D}_{\text{ref}}$. Consequently, the estimation is strictly bounded by the maximum reward observed in the static reference, denoted as $R_{\max} = \max_{y' \in \mathcal{D}_{\text{ref}}} R(y')$.

As the policy $\pi_{\theta}$ evolves, it inevitably explores high-reward regions and generates candidates that outperform the static reference (i.e., $R(y) > R_{\max}$). However, for any such superior candidate, the estimated quantile saturates at the upper bound: $F_{\text{ref}}(R(y)) = 1$. This leads to a critical failure in supervision: consider two generated lists $y_1$ and $y_2$ s.t. $R(y_1) > R(y_2) > R_{\max}$. Although $y_1$ is strictly superior, the static estimator assigns them identical quantile scores:
\begin{equation}
    (F_{\text{ref}}(R(y_1)))^{N-1} = (F_{\text{ref}}(R(y_2)))^{N-1} = 1.
\end{equation}
This indicates that the alignment process becomes \textbf{indiscriminate} in the high-utility region. By failing to distinguish between varying levels of superior quality, the objective effectively flattens the optimization landscape, preventing the policy from converging towards the globally optimal solution.

\paragraph{2. Gradient Decay}
We further analyze the gradient of the primary utility term in Eq.~(\ref{eq:implicit_reward}), the quantile reward. Let $J_{\text{qt}}(\theta) = \mathbb{E}_{y \sim \pi_\theta} [ (N-1) \log F_{\text{ref}}(R(y)) ]$. According to the policy gradient theorem, the gradient with respect to $\theta$ is:
\begin{equation}
    \nabla_\theta J_{\text{qt}}(\theta) = \mathbb{E}_{y \sim \pi_\theta} \left[ \nabla_\theta \log \pi_\theta(y) \cdot (N-1) \log F_{\text{ref}}(R(y)) \right].
\end{equation}
As the policy $\pi_\theta$ evolves, its probability mass shifts towards high-reward regions, increasingly overlapping with the sparse right tail of the static reference distribution. Consequently, even before the policy reaches the theoretical maximum, the generated responses $y$ fall into high-quantile regions where $F_{\text{ref}}(R(y)) \to 1$.
This leads to a rapid shrinkage of the effective weighting term:
\begin{equation}
    \lim_{F_{\text{ref}} \to 1} \log F_{\text{ref}}(R(y)) = 0.
\end{equation}
This implies the \textbf{gradient decay}, as the policy improves relative to the static reference, the gradient magnitude shrinks rapidly. This vanishing supervision signal leads to inefficient optimization, causing the training to stall prematurely before reaching the global optimum.

\subsection{Method Overview}
Building upon the theoretical analysis in \mysec{limitations}, we propose \textbf{BLADE}. As illustrated in Figure~\ref{fig:framework}, BLADE fundamentally shifts the alignment paradigm from anchoring to a fixed reference to pursuing a \textit{self-evolving target}.

From a Bayesian perspective, the standard static reference $\pi_{\text{ref}}$ serves as a prior belief about the reward landscape. While providing a stable starting point, this static prior imposes \textbf{Indiscriminate Supervision} on high-quality candidates discovered during the policy's exploration. To rectify this, BLADE reformulates the target construction as a dynamic estimation process. At each training step, we treat the real-time response lists generated by the evolving policy $\pi_{\theta}$ as \textbf{dynamic evidence}. By fusing this up-to-date evidence with the static prior, BLADE constructs a posterior target distribution that adaptively scales with the model's current capability.

This dynamic mechanism effectively resolves the identified bottlenecks. By incorporating real-time rollouts, BLADE restores the \textbf{ranking discrimination} for superior candidates, and by recalibrating the probability scale, it prevents \textbf{Gradient Decay}. \myfig{comparison_bon_blade} illustrates that while the static baseline stalls due to signal saturation, BLADE continuously updates the target to ensure the optimization gradients remain informative, enabling continuous evolution beyond the initial reference.

\subsection{Bayesian Dynamic Estimation}
\label{sec:dynamic_estimation}

To instantiate the self-evolving target distribution, the core challenge lies in robustly estimating the CDF of the rewards, $F(r) = P(R(\boldsymbol{y}) < r)$.
Instead of relying on a static approximation, BLADE casts this estimation as a real-time Bayesian inference problem.

\paragraph{Probabilistic Formulation.}
For each reward threshold $r$, we treat the corresponding quantile value $\theta_r = F(r)$ as a random variable. 
Since $\theta_r$ is bounded in $[0, 1]$, we model it using a threshold-specific \textbf{Beta distribution}:
\begin{equation}
    \theta_r \sim \text{Beta}(\alpha^r, \beta^r).
\end{equation}
The Beta family is a natural choice because its support matches the range of CDF values and it is conjugate to the binomial likelihood induced by counting how many samples fall below the threshold $r$, enabling efficient closed-form updates.

\paragraph{Bayesian Update Mechanism.}
We construct the estimator by fusing statistics from the static reference set $\mathcal{D}_{\text{ref}}$, comprising $M$ samples, and the current training batch $\mathcal{D}_{\text{batch}}$ of size $G$.
Let $N_{\text{ref}}^{<r}$ and $N_{\text{batch}}^{<r}$ denote the count of samples with rewards lower than $r$ in the respective sets.

We initialize the threshold-specific prior belief using the static reference statistics, where $\alpha_0^r = N_{\text{ref}}^{<r}$ and $\beta_0^r = M - N_{\text{ref}}^{<r}$.
To balance the stability of the static prior and the adaptivity of online rollouts, we use a \textbf{power-scaled likelihood} with a \textbf{dynamic coefficient} $\tau \ge 0$.
This coefficient controls the effective sample size of the dynamic evidence: $\tau=0$ recovers the static estimator, $\tau=1$ corresponds to a standard beta-binomial update using the current batch, and $0<\tau<1$ conservatively tempers the batch evidence to reduce estimation variance under limited rollouts.
Formally, the posterior is derived as:
\begin{equation}
\begin{aligned}
    P(\theta_r | \mathcal{D}_{\text{batch}}) &\propto P(\theta_r) \cdot \left[ P(\mathcal{D}_{\text{batch}} | \theta_r) \right]^\tau \\
    &\propto \underbrace{\theta_r^{\alpha_0^r - 1} (1-\theta_r)^{\beta_0^r - 1}}_{\text{Static Prior}} \cdot \left[ \underbrace{\theta_r^{N_{\text{batch}}^{<r}} (1-\theta_r)^{(G - N_{\text{batch}}^{<r})}}_{\text{Dynamic Evidence}} \right]^\tau \\
    &= \theta_r^{(\alpha_0^r + \tau N_{\text{batch}}^{<r}) - 1} (1-\theta_r)^{(\beta_0^r + \tau (G - N_{\text{batch}}^{<r})) - 1}.
\end{aligned}
\end{equation}
Identifying the exponents with the standard Beta distribution, we obtain the updated parameters for the posterior $\text{Beta}(\alpha_{\text{new}}^r, \beta_{\text{new}}^r)$:
\begin{equation}
    \alpha_{\text{new}}^r = \alpha_0^r + \tau \cdot N_{\text{batch}}^{<r}, \quad
    \beta_{\text{new}}^r = \beta_0^r + \tau \cdot (G - N_{\text{batch}}^{<r}).
\end{equation}

\paragraph{Posterior Estimation.}
Finally, to obtain a robust point estimate for the optimization target at threshold $r$, we take the expectation of the posterior distribution. This yields the \textbf{BLADE Dynamic Estimator}:
\begin{equation}
    \label{eq:blade_formula}
    \hat{F}_{\text{BLADE}}(r) 
    = \mathbb{E}[\theta_r | \mathcal{D}_{\text{batch}}] 
    = \frac{\alpha_{\text{new}}^r}{\alpha_{\text{new}}^r + \beta_{\text{new}}^r} 
    = \frac{N_{\text{ref}}^{<r} + \tau \cdot N_{\text{batch}}^{<r}}{M + \tau \cdot G}.
\end{equation}

\subsection{Efficient Optimization via Shared Sampling}
\label{sec:optimization}

To optimize the policy towards the dynamic target estimated in Eq.~\eqref{eq:blade_formula}, we employ Group Relative Policy Optimization (GRPO) as our optimization backbone. We select this algorithm specifically because its group-wise sampling mechanism naturally complements our Bayesian estimation strategy, enabling efficient, critic-free optimization.

\paragraph{Dynamic Reward Integration.}
Instead of relying on static reward signals, we construct the optimization objective dynamically using the estimator derived in the previous section. For each generated candidate list $y_i$ within the sampled group $\mathcal{D}_{\text{batch}}$, we formulate the proxy reward $r_i$ as:
\begin{equation}
    r_i = (N-1) \log \hat{F}_{\text{BLADE}}(R(y_i)).
    \label{eq:proxy_reward}
\end{equation}
This dynamic reward is then directly substituted into the GRPO objective to compute the group-relative advantage, ensuring that policy updates are consistently driven by the self-evolving target distribution.

\paragraph{Efficiency: The Shared Sampling Mechanism.}
A critical synergy in our framework is the \textbf{Shared Sampling} strategy. Both the Bayesian target estimation and the GRPO baseline calculation require a group of samples to derive distributional statistics. By utilizing the same batch of rollouts $\mathcal{D}_{\text{batch}}$ for a \textbf{dual purpose}, BLADE achieves \textit{zero-overhead} target estimation. Specifically, the generated rollouts are first utilized as "dynamic evidence" to update the Bayesian posterior, and immediately reused to compute the policy gradients. Consequently, BLADE maintains the same computational efficiency as static BoN alignment, achieving superior performance \textbf{without additional inference or training costs}.

%% file: sections/4.Experiments.tex
\section{Experiments}
\label{sec:experiments}
In this section, we conduct extensive experiments to evaluate BLADE by addressing the following research questions:
\begin{itemize}
    \item \textbf{RQ1 (Overall Performance):} Does BLADE achieve superior performance compared to SFT and state-of-the-art alignment baselines in list-wise recommendation tasks?
    
    \item \textbf{RQ2 (Mechanism Analysis):} How does the proposed Bayesian dynamic estimation mechanism contribute to the robustness and effectiveness of the alignment process?
    
    \item \textbf{RQ3 (Generalization):} Can BLADE serve as a general framework to effectively optimize diverse list-wise objectives beyond standard relevance?
\end{itemize}

\definecolor{colorLLM}{RGB}{242, 249, 255}   
\definecolor{colorDecode}{RGB}{248, 245, 255} 
\definecolor{colorList}{RGB}{245, 252, 245}   
\definecolor{colorOurs}{RGB}{255, 250, 240}   

\definecolor{lineGrey}{gray}{0.55} 
\newcolumntype{?}{!{\color{lineGrey}\vrule width 0.6pt}}

\subsection{Experimental Setup}

\subsubsection{Datasets}
We evaluate our method on three real-world datasets:
\emph{Amazon CDs and Vinyl}\footnote{\url{http://jmcauley.ucsd.edu/data/amazon/index_2014.html}}, 
\emph{Steam}\footnote{\url{https://cseweb.ucsd.edu/~jmcauley/datasets.html\#amazon_reviews}}, and 
\emph{Goodreads}\footnote{\url{https://mengtingwan.github.io/data/goodreads}}, with detailed statistics provided in \mytab{dataset_stats}. Following~\cite{bao2023tallrec, gao2025sprec}, we filtered out users with fewer than 10 interactions. For the Steam and Goodreads datasets, we retained the top 15\% most popular items to ensure dense interaction history. For list-wise sequence construction, we applied a sliding window approach, utilizing the past 10 items as the user history context and the subsequent 10 items as the target recommendation list. Regarding the training setup, we adhered to a standard two-stage paradigm: the base model was first supervised fine-tuned on the full dataset to acquire domain knowledge; subsequently, to balance computational costs while leveraging the data efficiency of preference alignment, the alignment phase was conducted on a sampled subset of 4,096 training and 512 validation instances. Crucially, to ensure unbiased evaluation, all results are reported on the complete test set.


\begin{table}[htbp]
    \centering
    \caption{Summary of dataset statistics.}
    \label{tab:dataset_stats}
    \begin{tabular}{lrrr}
        \toprule
        \textbf{Dataset} & \textbf{\# Users} & \textbf{\# Items} & \textbf{\# Interactions} \\
        \midrule
        Steam & 3,050 & 1,473 & 23,217 \\
        Goodreads & 5,762 & 3,261 & 101,292 \\
        Amazon CDs & 21,347 & 13,078 & 185,855 \\
        \bottomrule
    \end{tabular}
\end{table}

\subsubsection{Baselines}
We compare BLADE against the following baselines:
\begin{itemize}
    \item \textbf{BIGRec:}~\cite{bao2023bi} A standard SFT baseline that fine-tunes LLMs for recommendation tasks, serving as the foundation for subsequent alignment methods.
    \item \textbf{S-DPO:}~\cite{chen2024softmax} An adaptation of DPO for recommendation by introducing a softmax loss over multiple negative samples.
    \item \textbf{ReRe:}~\cite{tan2025reinforced} An RL-based paradigm utilizing constrained beam search for efficient sampling and auxiliary ranking rewards for fine-grained supervision.
    \item \textbf{SPRec:}~\cite{gao2025sprec} A self-play framework that treats model generations as negative samples to achieve self-evolution and debiasing.
    \item \textbf{Beam Search:}~\cite{sutskever2014sequence} A representative point-wise decoding strategy that selects the top-$K$ independent hypotheses based solely on sequence probability.
    \item \textbf{D$^3$:}~\cite{bao2024decoding} A decoding approach that mitigates amplification bias and homogeneity by calibrating length normalization and penalizing repetitive generations.
    \item \textbf{List DPO:}~\cite{rafailov2024direct} Follows the standard RLHF protocol to sample candidate list pairs and optimizes them via DPO based on relative rewards.
    \item \textbf{BoN Alignment:}~\cite{amini2024variational} The static alignment strategy described in \mysec{bon_framework}, which distills Best-of-N capabilities using a fixed reference.
\end{itemize}

\begin{table*}[t]
\centering
\caption{
    Overall performance comparison. 
    The background colors distinguish different categories of methods: 
    \colorbox{colorLLM}{LLM-based}, 
    \colorbox{colorDecode}{Inference Strategies}, 
    \colorbox{colorList}{List-wise}, 
    and \colorbox{colorOurs}{Ours}. 
    \textbf{R@\textit{k}} and \textbf{N@\textit{k}} denote Recall@\textit{k} and NDCG@\textit{k}, respectively. 
    Specific to our methods, \textbf{BLADE-R} and \textbf{BLADE-N} utilize Recall and NDCG as the optimization reward signal, respectively.
    The best results are highlighted in \textbf{bold}, and the second-best results are \underline{underlined}.
}
\label{tab:main_results_tuned}

\setlength{\tabcolsep}{6pt} 

\resizebox{\textwidth}{!}{%
\begin{tabular}{l cccc ? cccc ? cccc}
\toprule
\multirow{2}{*}{\textbf{Method}} 
 & \multicolumn{4}{c?}{\textbf{Amazon CDs and Vinyl}} 
 & \multicolumn{4}{c?}{\textbf{Steam}} 
 & \multicolumn{4}{c}{\textbf{Goodreads}} \\
\cmidrule(lr){2-5} \cmidrule(lr){6-9} \cmidrule(lr){10-13}
 & \textbf{R@3} & \textbf{N@3} & \textbf{R@5} & \textbf{N@5} 
 & \textbf{R@3} & \textbf{N@3} & \textbf{R@5} & \textbf{N@5} 
 & \textbf{R@3} & \textbf{N@3} & \textbf{R@5} & \textbf{N@5} \\
\midrule

\rowcolor{colorLLM}
BIGRec   & 0.0079 & 0.0331 & 0.0103 & 0.0283 & 0.0101 & 0.0375 & 0.0158 & 0.0357 & 0.0166 & 0.0602 & 0.0195 & 0.0545 \\
\rowcolor{colorLLM}
S-DPO     & 0.0099 & 0.0391 & 0.0120 & 0.0319 & 0.0093 & 0.0329 & 0.0147 & 0.0317 & 0.0175 & 0.0642 & 0.0195 & 0.0605 \\
\rowcolor{colorLLM}
ReRe     & 0.0110 & 0.0410 & 0.0135 & 0.0345 & 0.0106 & 0.0382 & 0.0157 & 0.0358 & 0.0176 & 0.0644 & 0.0202 & 0.0584 \\
\rowcolor{colorLLM}
SPRec    & 0.0104 & 0.0409 & 0.0125 & 0.0336 & 0.0109 & 0.0397 & \underline{0.0168} & 0.0374 & 0.0175 & 0.0639 & 0.0195 & 0.0606 \\

\addlinespace 

\rowcolor{colorDecode}
Beam Search & 0.0069 & 0.0251 & 0.0075 & 0.0190 & 0.0070 & 0.0253 & 0.0073 & 0.0187 & 0.0013 & 0.0039 & 0.0013 & 0.0028 \\
\rowcolor{colorDecode}
D$^3$       & 0.0094 & 0.0345 & 0.0119 & 0.0284 & 0.0069 & 0.0232 & 0.0107 & 0.0221 & 0.0118 & 0.0409 & 0.0163 & 0.0358 \\

\addlinespace 

\rowcolor{colorList}
List DPO & 0.0095 & 0.0380 & 0.0121 & 0.0308 & 0.0096 & 0.0336 & 0.0151 & 0.0325 & 0.0177 & 0.0649 & 0.0196 & \underline{0.0612} \\
\rowcolor{colorList}
BoN Alignment    & 0.0098 & 0.0401 & 0.0120 & 0.0333 & \underline{0.0114} & \underline{0.0416} & 0.0164 & 0.0384 & 0.0179 & 0.0653 & \underline{0.0212} & 0.0588 \\

\addlinespace

\rowcolor{colorOurs}
BLADE-R  & \textbf{0.0130} & \underline{0.0451} & \textbf{0.0156} & \underline{0.0379} 
         & \textbf{0.0116} & \textbf{0.0419} & \textbf{0.0171} & \textbf{0.0395} 
         & \textbf{0.0184} & \underline{0.0672} & \textbf{0.0219} & 0.0606 \\
\rowcolor{colorOurs}
BLADE-N  & \underline{0.0119} & \textbf{0.0474} & \underline{0.0144} & \textbf{0.0410} 
         & 0.0111 & 0.0407 & 0.0155 & \underline{0.0393} 
         & \underline{0.0183} & \textbf{0.0686} & 0.0206 & \textbf{0.0618} \\

\bottomrule
\end{tabular}
}
\end{table*}

\subsubsection{Evaluation Metrics}
\label{sec:metrics}

We employ two widely adopted metrics to evaluate recommendation accuracy: \textbf{NDCG@k} and \textbf{Recall@k}, reporting performance with $k \in \{3, 5\}$. NDCG assesses the ranking quality by assigning higher scores to relevant items at top positions, while Recall measures the proportion of relevant items successfully retrieved. Additionally, to assess the comprehensive alignment capabilities (RQ3), we utilize:

\noindent \textbf{Mean Group Unfairness (MGU)}~\cite{jiang2024item}. 
MGU measures the disparity between the generated list $L$ and user history $H$. It calculates the Mean Absolute Error (MAE) of genre distributions over the genre set $\mathcal{C}$:
\begin{equation}
    \text{MGU} = \frac{1}{|\mathcal{C}|} \sum_{c \in \mathcal{C}} |P(c|L) - P(c|H)|,
\end{equation}
where $P(c|\cdot)$ denotes the normalized frequency of genre $c$.

\noindent \textbf{Intra-List Diversity (ILD)} ~\cite{Ziegler2005ImprovingRL}. 
ILD evaluates diversity by computing the average pairwise Jaccard distance between items in list $L$ of size $N$. Let $g_i$ be the genre set of item $i$:
\begin{equation}
    \text{ILD} = \frac{1}{N(N-1)} \sum_{i \neq j} \left( 1 - \frac{|g_i \cap g_j|}{|g_i \cup g_j|} \right).
\end{equation}
A higher ILD indicates lower redundancy among recommended items.

\subsubsection{Implementation Details}
\label{sec:implementation}

We utilize \textbf{Llama-3.2-1B-Instruct}~\cite{grattafiori2024llama} as the backbone and conduct all experiments on 4 NVIDIA A100 GPUs. The policy is first initialized via standard SFT on the full dataset for 3 epochs with a learning rate of $1\times 10^{-4}$. For BLADE optimization, we train for 3 epochs using the AdamW optimizer with a learning rate of $5\times 10^{-6}$ and a linear decay scheduler. We set the group size to $G=16$ and the reference set size to $M=128$. The dynamic coefficient $\tau$ is tuned within the set $\{0.1, 0.3, 0.5, 1.0\}$. During the training phase, we employ a sampling strategy with a temperature of 0.8 and a repetition penalty of 1.05 to encourage exploration.

Regarding baselines, we generally follow the default configurations of the original papers. Notably, for DPO-based methods (S-DPO, List-DPO), we perform a warm-up SFT on the positive samples of the alignment subset prior to preference optimization. This ensures adequate distribution coverage and training stability. For DPO training, the KL penalty $\beta$ is set to 0.1, and for S-DPO, we utilize 3 negative samples per positive instance. Crucially, to ensure reproducibility, evaluations for all training-based methods are performed using greedy decoding. For decoding-specific baselines (i.e., Beam Search and $D^3$), we adhere to their standard protocols.

\subsection{Overall Performance}

\mytab{main_results_tuned} presents the overall performance comparison of BLADE against state-of-the-art baselines across three real-world datasets. The results demonstrate that \textbf{BLADE consistently outperforms all competitive methods}, achieving superior scores on the majority of metrics across all datasets. Our analysis yields the following key observations:

\paragraph{Breaking the Limits of Static BoN Alignment.}
The most significant finding is that BLADE substantially surpasses existing list-wise approaches. On the \textbf{Amazon CDs} dataset, \textbf{BLADE-N} achieves a remarkable \textbf{23.1\%} improvement in NDCG@5 and an \textbf{18.2\%} improvement in NDCG@3 compared to the strongest list-wise baseline, BoN Alignment. This empirical evidence strongly validates the effectiveness of our approach in overcoming the performance upper bound of static reference models, unlocking the potential to discover high-quality ranking lists that static methods fail to capture.
\begin{figure}[t]
    \centering
    \begin{subfigure}{0.48\linewidth}
        \centering
        \includegraphics[width=\linewidth]{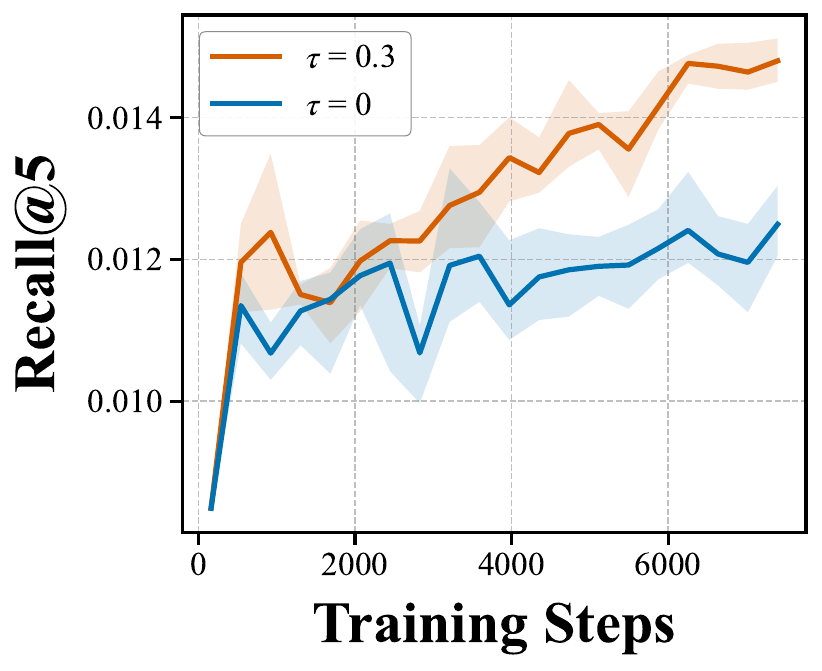} 
        \caption{Training Recall curves}
        \label{fig:train_process}
    \end{subfigure}
    \hfill 
    \begin{subfigure}{0.48\linewidth}
        \centering
        \includegraphics[width=\linewidth]{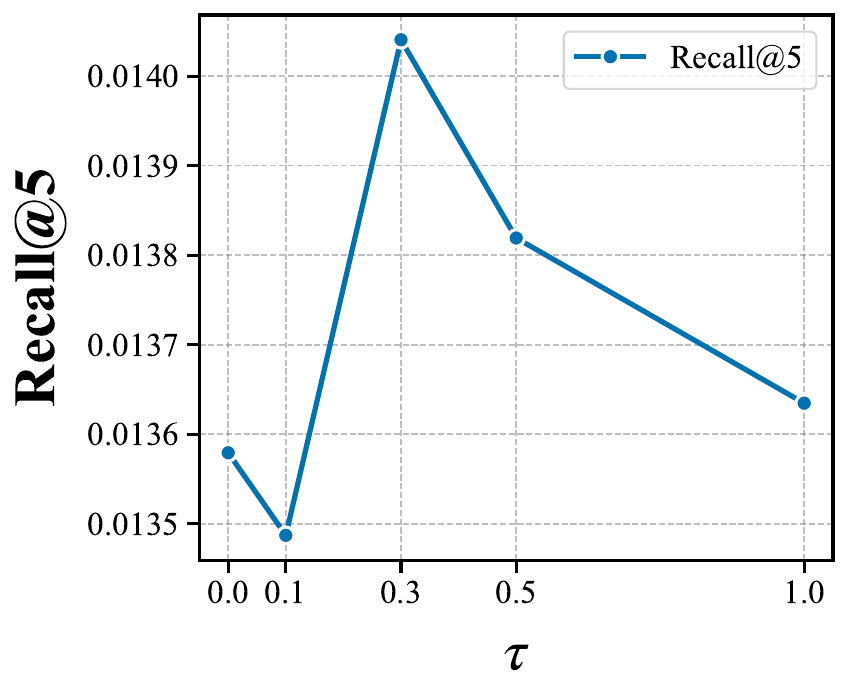}
        \caption{Impact of $\tau$}
        \label{fig:tau_impact}
    \end{subfigure}
    
    \caption{
        \textbf{Study on CDs and Vinyl datasets.} 
        \subref{fig:train_process} Comparison of Recall evolution during training with $\tau=0$ and $\tau=0.3$. 
        \subref{fig:tau_impact} Performance sensitivity of Test Recall with respect to varying $\tau$.
    }
    \label{fig:tau_analysis}
\end{figure}
\paragraph{Superiority over Standard LLM-based Methods.}
Compared to standard LLM-based alignment baselines, BLADE demonstrates a clear performance advantage. While methods like S-DPO and BIGRec show improvements over the base SFT model, they still fail to match BLADE's capability in optimizing holistic list-level utility. This indicates that directly optimizing global list-wise objectives via our Bayesian framework is significantly more effective than optimizing the local preference signals utilized in standard alignment techniques.

\paragraph{Inadequacy of Inference-time Strategies.}
While inference-time strategies attempt to optimize recommendation lists without parameter updates, our results reveal their fundamental limitations. 
The suboptimal performance of Beam Search is directly attributed to its inherent independence assumption. We empirically observe that independently sampling multiple items often leads to severe semantic repetition, as the model greedily collapses into a narrow distribution of high-probability candidates. 
Although D$^3$ alleviates this homogeneity through explicit decoding penalties, it fundamentally fails to account for the complex relationships between items. 
Such heuristic constraints merely suppress repetition without capturing the holistic coherence or compatibility required for high-quality lists. 
In contrast, BLADE internalizes these complex inter-item dependencies directly into the model parameters, naturally generating diverse and coherent recommendations.

\begin{figure}
    \centering
    \includegraphics[width=\linewidth]{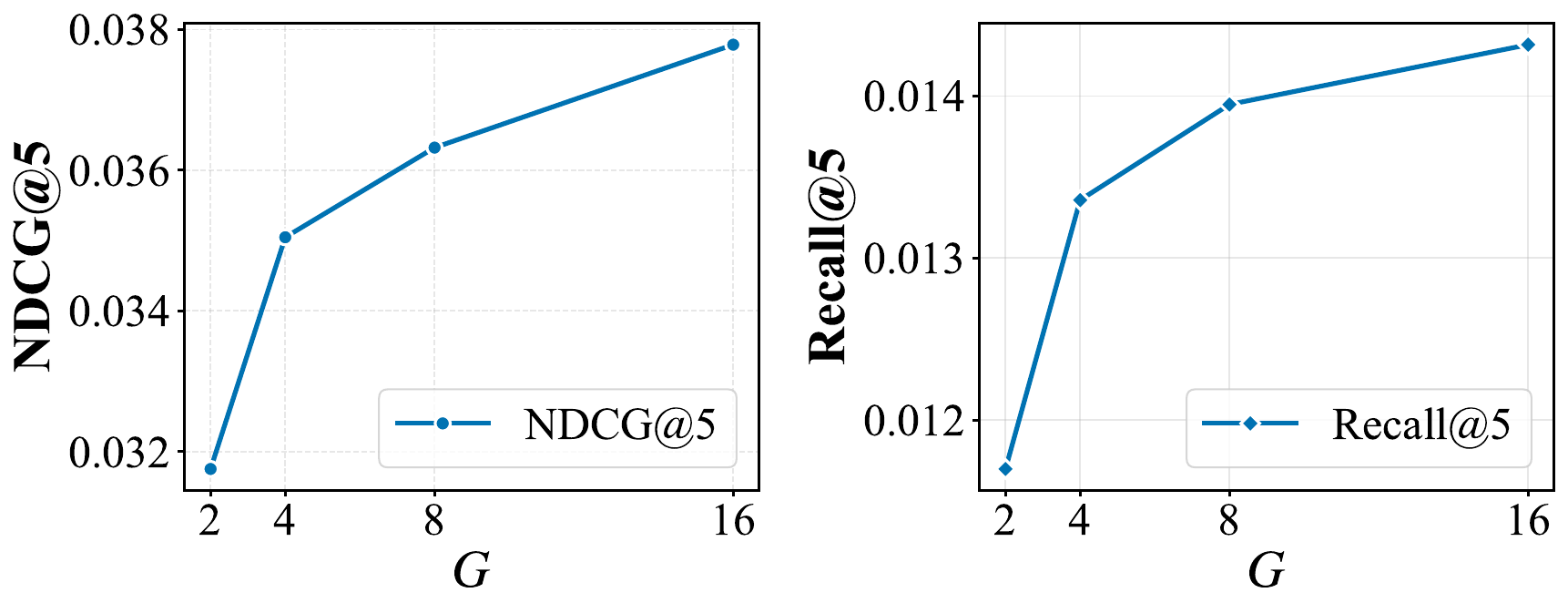}
    \caption{Effect of the number of generated items (G) in BLADE}
    \label{fig:G_scaling}
\end{figure}

\paragraph{Optimization Flexibility and Controllability.}
A unique advantage of our framework is its ability to align with specific user-defined objectives. As evidenced by the performance trade-off between BLADE-R and BLADE-N, the model effectively adapts to the chosen optimization target. Specifically, on the Goodreads and Amazon CDs datasets, BLADE-R excels in Recall metrics, whereas BLADE-N consistently achieves the highest scores on NDCG metrics. This distinction highlights the \textbf{controllability} of BLADE: it explicitly steers the policy towards the specific reward landscape required by the downstream recommendation task.

\subsection{Analysis of Bayesian Dynamic Estimation}
To empirically validate our proposed framework, we conduct an in-depth analysis of the Bayesian dynamic estimation mechanism. Specifically, we investigate how the dynamic coefficient $\tau$ mitigates the gradient decay problem and how the Bayesian prior \textbf{compensates for sampling sparsity} under varying inference budgets.

\subsubsection{Impact of Dynamic Coefficient $\tau$}
We first verify the critical role of the dynamic coefficient $\tau$ in sustaining optimization progress. \myfig{train_process} illustrates the Recall curves during training, while \myfig{tau_impact} presents the performance sensitivity with respect to $\tau$.
\paragraph{Mitigating Gradient Decay.}
As shown in Figure~\ref{fig:tau_analysis}(a), the training dynamics of the static baseline ($\tau=0$) and BLADE ($\tau=0.3$) diverge significantly. The static baseline improves initially but rapidly plateaus. This stagnation empirically validates our hypothesis regarding \textit{Gradient Decay}: as the model improves, the fixed reference distribution fails to provide informative signals, causing the training to stall prematurely. In contrast, BLADE sustains a steady improvement, confirming that updating the target distribution dynamically ensures effective learning throughout the process.

\begin{figure}
    \centering
    \includegraphics[width=1\linewidth]{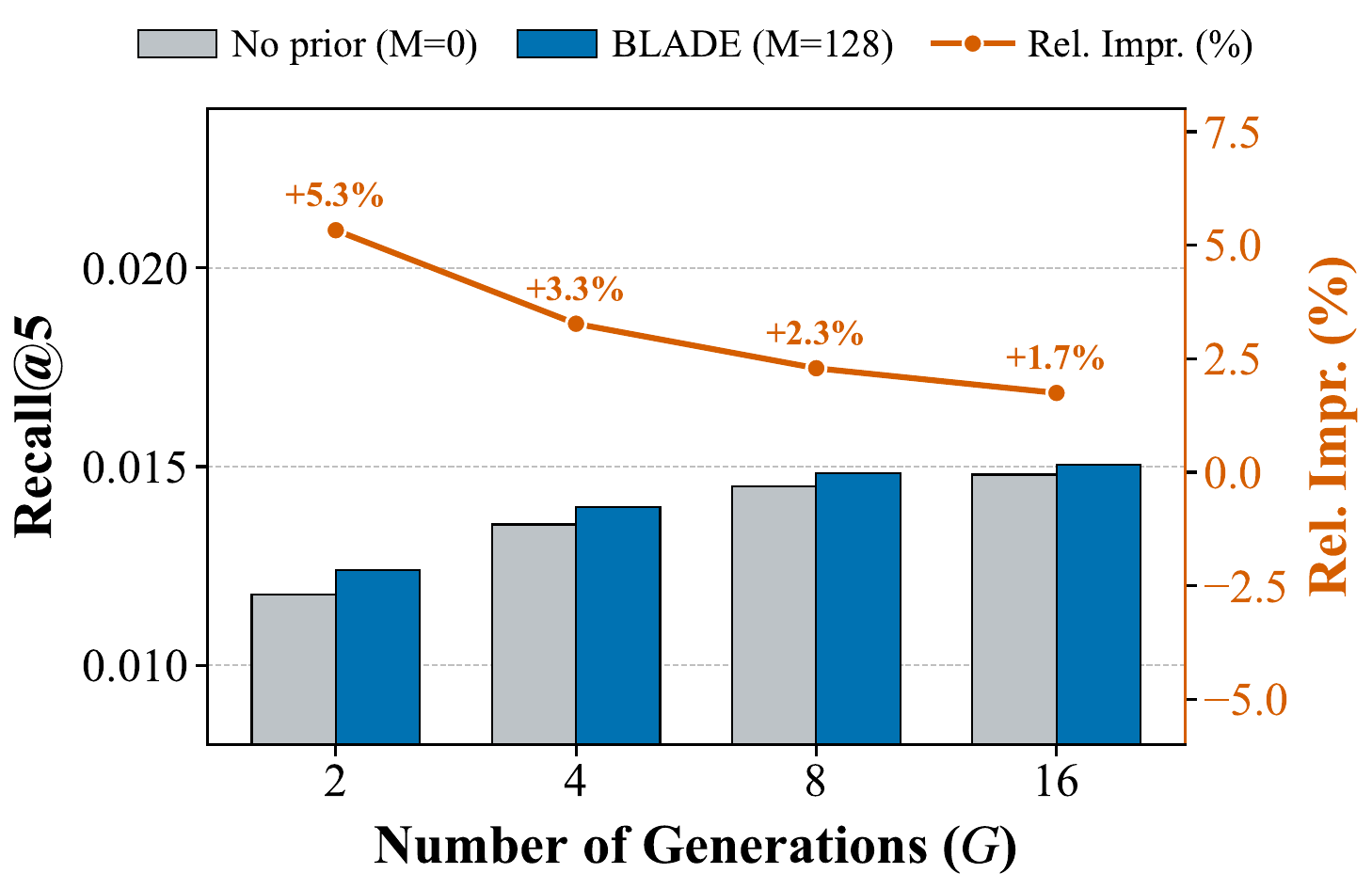}
    \caption{Robustness and data efficiency analysis across varying generation sizes ($G$). The bars (left axis) compare the absolute Recall@5 performance. The overlaid line (right axis) tracks the relative improvement, highlighting substantial gains in low-resource regimes.}
    \label{fig:robust}
\end{figure}
\paragraph{Balancing Prior and Evidence.}
Figure~\ref{fig:tau_analysis}(b) further illustrates the effect of $\tau$. We observe that the performance initially improves as $\tau$ increases but subsequently degrades when $\tau$ becomes too large. A small $\tau$ over-anchors the model to the suboptimal static prior, restricting exploration. Conversely, a large $\tau$ introduces high variance due to the limited batch samples. This indicates that a moderate $\tau$ strikes an optimal balance, leveraging the stability of the static prior while effectively adapting to the model's growing capabilities.

\subsubsection{Robustness and Estimation Stability}

\paragraph{Scaling with Compute.}
First, as shown in \myfig{G_scaling}, we observe that both NDCG@5 and Recall@5 improve consistently as $G$ increases from 2 to 16. This confirms that our method effectively benefits from a larger search space during training. This scalability implies that the model can achieve further alignment gains when more computational resources are available.

\paragraph{Stabilizing Role of the Prior.}
More importantly, we highlight the stabilizing role of the Bayesian prior. As shown in \myfig{robust}, BLADE consistently outperforms the "No Prior" baseline. The relative improvement is most significant when the group size is small. In this scenario, the dynamic evidence from the current batch is unstable due to sampling sparsity, and the static prior acts as a stabilizer, preventing estimation biases. As $G$ increases, the gap between the two methods narrows because the dynamic evidence becomes more reliable. This demonstrates that BLADE is highly robust: it yields substantial improvements over the baseline even with a minimal sampling cost, making it suitable for resource-constrained applications.

\subsection{Generalizability to Diverse List-wise Targets}
\label{sec:generalizability}

\begin{figure}[t]
    \centering
    \begin{subfigure}[b]{0.48\linewidth}
        \centering
        \includegraphics[width=\linewidth]{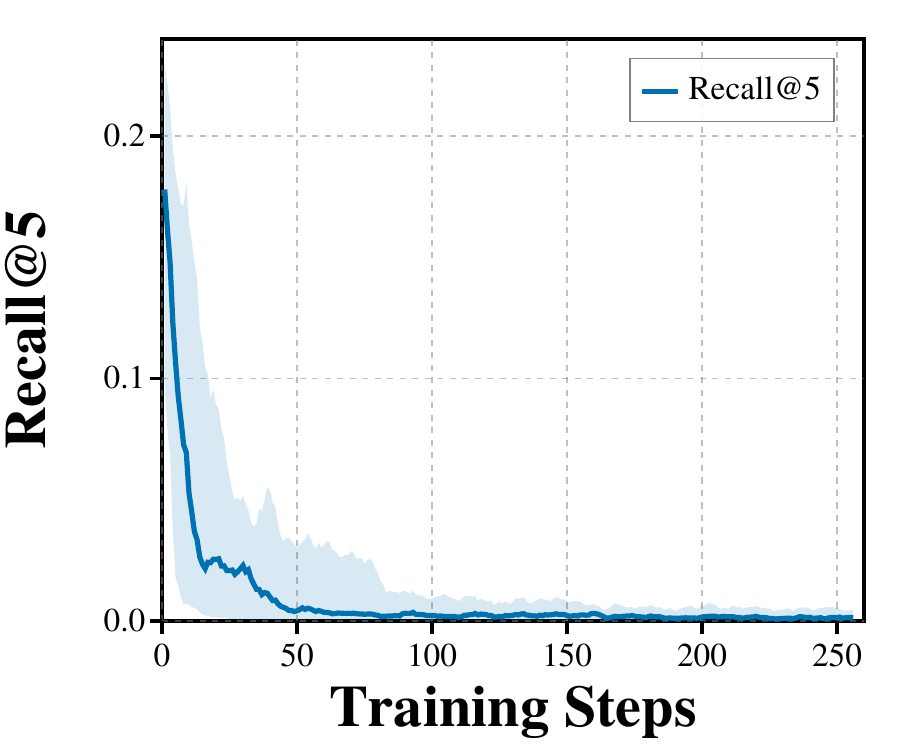}
        \caption{Recall@5}
        \label{fig:aux_recall_curve}
    \end{subfigure}
    \hfill
    \begin{subfigure}[b]{0.48\linewidth}
        \centering
        \includegraphics[width=\linewidth]{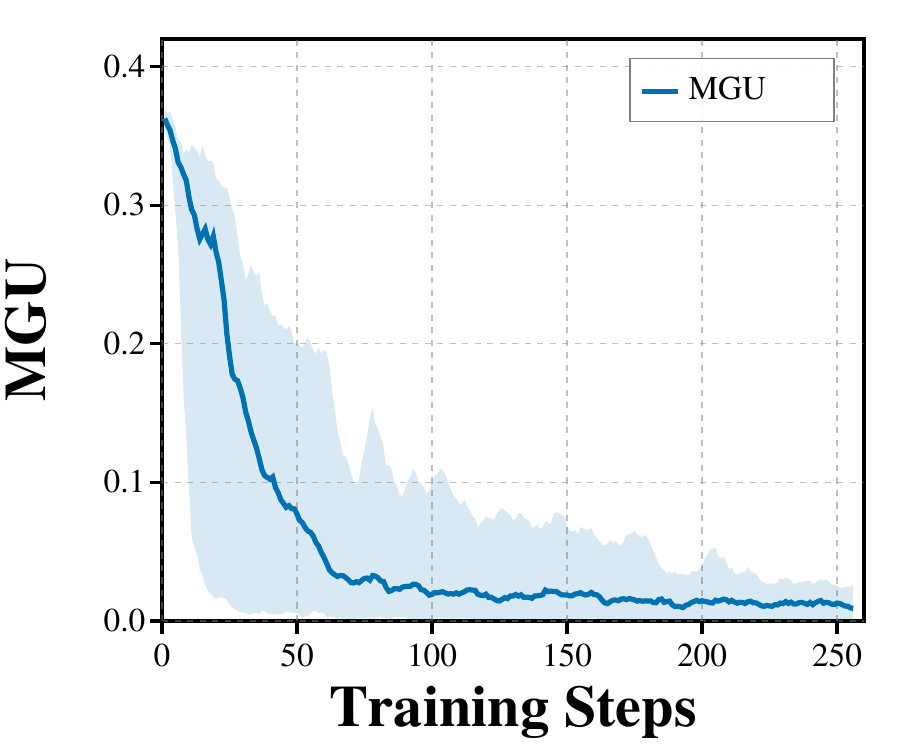}
        \caption{MGU}
        \label{fig:aux_mgu_curve}
    \end{subfigure}
    \caption{
    Training diagnostics under standalone fairness-objective optimization.
    Recall@5 rapidly decays toward zero, while MGU also decreases during training.
    This indicates that directly optimizing the auxiliary fairness objective can improve the target list-wise metric but severely degrades recommendation relevance.
    }
    \label{fig:aux_diagnostic}
\end{figure}

\begin{figure}[t]
    \centering
    \begin{subfigure}[b]{0.8\linewidth}
        \centering
        \includegraphics[width=\linewidth]{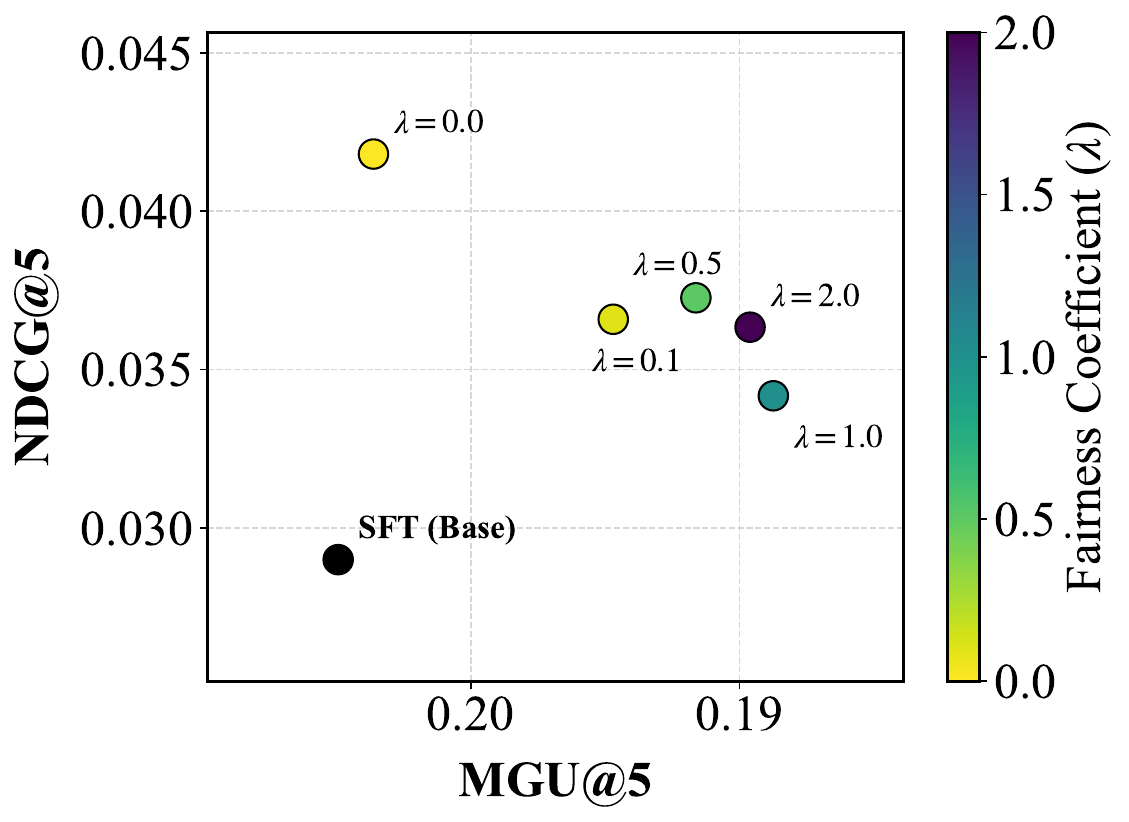}
        \caption{Fairness vs. Accuracy}
        \label{fig:fairness_pareto}
    \end{subfigure}
    \hfill 
    \begin{subfigure}[b]{0.8\linewidth}
        \centering
        \includegraphics[width=\linewidth]{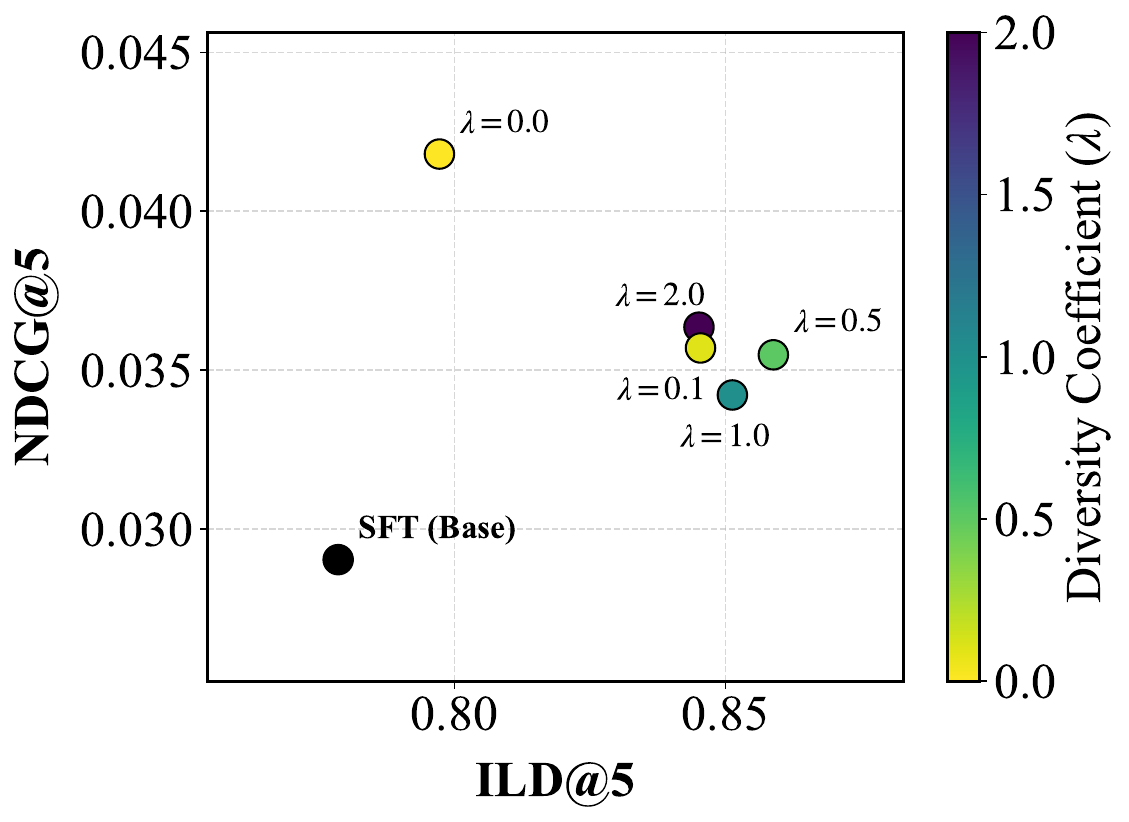}
        \caption{Diversity vs. Accuracy}
        \label{fig:diversity_pareto}
    \end{subfigure}
    
    \caption{Performance comparison on list-wise metrics. (a) Fairness (MGU, lower is better) vs. Accuracy. (b) Diversity vs. Accuracy. Both figures demonstrate BLADE's versatility, achieving substantial improvements in the respective list-wise metrics while simultaneously maintaining superior accuracy over the SFT baseline.}
    \label{fig:rq3_pareto}
\end{figure}
In this section, we investigate the generalizability of BLADE as a metric-agnostic alignment framework. Specifically, we evaluate its performance on two distinct objectives: \textbf{MGU} and \textbf{ILD}.

\paragraph{Composite Reward Formulation.}
Empirically, we found that optimizing the policy solely with auxiliary list-wise objectives can lead to rapid degradation in recommendation accuracy.
These objectives capture distributional properties of the generated list, but do not by themselves provide a relevance-grounding signal for recommendation.
As shown in Figure~\ref{fig:aux_diagnostic}, standalone fairness-objective optimization reduces MGU during training, but the Recall@5 signal quickly decays toward zero.
This indicates that the policy can improve the auxiliary objective while drifting away from relevant recommendation behavior.
Consequently, we formulate a \textbf{composite reward} by incorporating the relevance score $R_{\text{NDCG}}$ as a grounding signal to maintain utility. The rewards are defined as:
\begin{align}
    R_{\text{fair}}(\boldsymbol{y}) &= R_{\text{NDCG}}(\boldsymbol{y}) - \lambda \cdot \text{MGU}(\boldsymbol{y}), \\
    R_{\text{div}}(\boldsymbol{y}) &= R_{\text{NDCG}}(\boldsymbol{y}) + \lambda \cdot \text{ILD}(\boldsymbol{y}),
\end{align}
where $\lambda$ determines the relative weight of the fairness/diversity reward.

\paragraph{Optimization of Fairness.}
We evaluate the model's ability to mitigate unfairness using Mean Group Unfairness (MGU). The results in Figure~\ref{fig:fairness_pareto} reveal two key findings:
\begin{enumerate}
    \item \textbf{Pareto Dominance:} The SFT baseline resides in a suboptimal region characterized by low accuracy and high unfairness. In contrast, BLADE achieves strict Pareto dominance: all experimental points, regardless of $\lambda$, are located to the top-right of the baseline. This demonstrates that BLADE identifies a superior solution space where accuracy and fairness coexist, rather than strictly trading one for the other.
    \item \textbf{Non-monotonic Sensitivity:} We observe that the improvement in fairness is not strictly monotonic with respect to $\lambda$. While introducing $\lambda$ significantly reduces MGU compared to the unconstrained case ($\lambda=0$), increasing $\lambda$ beyond a certain threshold yields diminishing returns in fairness while incurring a cost in NDCG. This suggests the existence of a ``sweet spot'' (e.g., $\lambda \approx 0.5$) in the composite reward landscape.
\end{enumerate}
\paragraph{Optimization of Intra-List Diversity.}
We further assess the model's capability to optimize Diversity. As shown in Figure~\ref{fig:diversity_pareto}:
\begin{enumerate}
    \item \textbf{Significant Diversity Gain:} Both SFT and the unconstrained BLADE ($\lambda=0$) exhibit low diversity, tending to recommend redundant popular items. However, introducing even a minimal coefficient ($\lambda=0.1$) triggers a \textbf{rapid adaptation}, boosting the ILD score significantly. This confirms that BLADE is highly responsive to sparse high-diversity signals that are typically ignored by static baselines.
    \item \textbf{High-Utility Diversity:} Crucially, even at the peak diversity point ($\lambda=0.5$), BLADE maintains an NDCG significantly higher than the SFT baseline. This indicates that the diversity gain is not achieved through random exploration, but by successfully retrieving high-quality items that are both relevant and diverse.
\end{enumerate}
In summary, these empirical findings firmly establish BLADE as a robust, metric-agnostic alignment framework capable of transcending the limitations of single-objective optimization. Unlike specialized methods tailored for narrow tasks, BLADE demonstrates exceptional versatility in handling complex, non-differentiable list-wise metrics through a unified Bayesian approach. 

%% file: sections/5.Related.tex
\section{Related work}
\label{sec:related}

In this section, we first briefly review LLM-based recommender systems and the challenges associated with list-wise optimization. Subsequently, we introduce the Best-of-N sampling method and its fine-tuning variants.

\subsection{LLMs for Recommendation}
The integration of Large Language Models (LLMs) into recommender systems has created a new paradigm, leveraging their extensive world knowledge and semantic reasoning capabilities. Existing research can be broadly categorized into three architectures: 
(1) \textbf{Feature Encoders}~\cite{ren2024representation,wang2024learnable,chen2024hllm}, which utilize the internal representations of LLMs to enhance the semantic embeddings of traditional collaborative filtering models; 
(2) \textbf{User Modeling}~\cite{zhang2024agentcf,liu2025onerec}, where LLMs serve as user profilers to infer latent interests or generate natural language explanations for recommendations; and 
(3) \textbf{Generative Recommendation}~\cite{kong2025minionerec,LLaRA,gao2025sprec,lin2025order}, the focus of this work, which formulates recommendation as a sequence-to-sequence generation task. In this paradigm, the model is prompted with user history and directly outputs item titles or identifiers.

Despite the promise of Generative Recommendation, pre-trained LLMs often suffer from domain shift and lack the specific collaborative filtering signals required for accurate ranking~\cite{bao2023tallrec}. While Supervised Fine-Tuning (SFT)~\cite{bao2023bi,P5} and Direct Preference Optimization (DPO)~\cite{liao2024rosepo,chen2024softmax} have been widely adopted to align LLMs with recommendation data, they typically rely on token-level cross-entropy or pair-wise ranking objectives. These local objectives are often misaligned with the holistic, list-wise metrics (e.g., NDCG, Diversity) that define user satisfaction in real-world scenarios, creating a gap between training objectives and inference-time performance.

\subsection{List-wise Recommendation}
Standard point-wise recommender systems learn individual preferences~\cite{he2017neural,kang2018self}, misaligning with real-world scenarios where users are exposed to complete item lists~\cite{cao2007learning,ai2018learning}. Consequently, user satisfaction relies on holistic list-level metrics (e.g., NDCG) that capture global item interactions. However, optimizing these metrics is fundamentally challenging because they are inherently complex, discrete, and non-differentiable, making them difficult to address with standard gradient-based objectives.

To address this challenge, industrial systems often employ a ``Generator-Evaluator'' pipeline for re-ranking~\cite{zhang2025generation,yang2025comprehensive}, which is difficult to unify into the end-to-end generative paradigm. Existing approaches utilize differentiable surrogate losses to approximate discrete ranking metrics~\cite{yang2025breaking,chao2024make}. However, such approximation inevitably suffers from estimation bias and poor generalization, as the smooth surrogate objectives diverge from the actual, non-differentiable evaluation protocols. Alternatively, Direct Reinforcement Learning theoretically supports non-differentiable rewards but proves notoriously unstable in practice, exhibiting high variance and sensitivity to reward scaling~\cite{chen2019top}. To overcome these limitations, we propose BLADE, which bypasses gradient constraints by internalizing list-wise search capabilities directly into the model weights.

\subsection{Best-of-N Sampling and Alignment}

Best-of-N (BoN) is a robust inference-time strategy proposed to optimize language generation quality by selecting the best response from $N$ candidates~\cite{stiennon2020learning}. Despite its simplicity, BoN has been proven to rival or even surpass complex alignment methods~\cite{gui2024bonbon}. However, the requirement for $N$ times the inference latency renders it computationally inefficient for practical applications.

To address this efficiency bottleneck, \textbf{BoN Alignment} aims to distill the search capability into the model weights. Early approaches employ \textit{offline distillation} by fine-tuning the model on the best samples generated by a teacher policy~\cite{dong2023raft}. More recent methods formulate BoN matching as a constrained optimization problem, solving it via DPO on best-worst pairs or online Reinforcement Learning\cite{gui2024bonbon,sessa2024bond,amini2024variational}.  While the latter achieves a better balance between maximizing reward and maintaining proximity to the reference model, we argue that it still suffers from Indiscriminate Supervision, and Gradient Decay, limiting their effectiveness in recommendation. Therefore, we propose BLADE, which leverages dynamic estimation of the real-time BoN distribution. This approach surpasses the static performance upper bound without additional training costs, achieving superior optimization for list-wise recommendation.

%% file: BLADE.bbl

\begin{thebibliography}{45}


\ifx \showCODEN    \undefined \def \showCODEN     #1{\unskip}     \fi
\ifx \showISBNx    \undefined \def \showISBNx     #1{\unskip}     \fi
\ifx \showISBNxiii \undefined \def \showISBNxiii  #1{\unskip}     \fi
\ifx \showISSN     \undefined \def \showISSN      #1{\unskip}     \fi
\ifx \showLCCN     \undefined \def \showLCCN      #1{\unskip}     \fi
\ifx \shownote     \undefined \def \shownote      #1{#1}          \fi
\ifx \showarticletitle \undefined \def \showarticletitle #1{#1}   \fi
\ifx \showURL      \undefined \def \showURL       {\relax}        \fi
\providecommand\bibfield[2]{#2}
\providecommand\bibinfo[2]{#2}
\providecommand\natexlab[1]{#1}
\providecommand\showeprint[2][]{arXiv:#2}

\bibitem[Ai et~al\mbox{.}(2018)]%
        {ai2018learning}
\bibfield{author}{\bibinfo{person}{Qingyao Ai}, \bibinfo{person}{Keping Bi}, \bibinfo{person}{Jiafeng Guo}, {and} \bibinfo{person}{W~Bruce Croft}.} \bibinfo{year}{2018}\natexlab{}.
\newblock \showarticletitle{Learning a deep listwise context model for ranking refinement}. In \bibinfo{booktitle}{\emph{The 41st international ACM SIGIR conference on research \& development in information retrieval}}. \bibinfo{pages}{135--144}.
\newblock


\bibitem[Amini et~al\mbox{.}(2024)]%
        {amini2024variational}
\bibfield{author}{\bibinfo{person}{Afra Amini}, \bibinfo{person}{Tim Vieira}, \bibinfo{person}{Elliott Ash}, {and} \bibinfo{person}{Ryan Cotterell}.} \bibinfo{year}{2024}\natexlab{}.
\newblock \showarticletitle{Variational best-of-n alignment}.
\newblock \bibinfo{journal}{\emph{arXiv preprint arXiv:2407.06057}} (\bibinfo{year}{2024}).
\newblock


\bibitem[Bao et~al\mbox{.}(2025)]%
        {bao2023bi}
\bibfield{author}{\bibinfo{person}{Keqin Bao}, \bibinfo{person}{Jizhi Zhang}, \bibinfo{person}{Wenjie Wang}, \bibinfo{person}{Yang Zhang}, \bibinfo{person}{Zhengyi Yang}, \bibinfo{person}{Yanchen Luo}, \bibinfo{person}{Chong Chen}, \bibinfo{person}{Fuli Feng}, {and} \bibinfo{person}{Qi Tian}.} \bibinfo{year}{2025}\natexlab{}.
\newblock \showarticletitle{A Bi-Step Grounding Paradigm for Large Language Models in Recommendation Systems}.
\newblock \bibinfo{journal}{\emph{ACM Transactions on Recommender Systems (TORS)}} (\bibinfo{year}{2025}).
\newblock


\bibitem[Bao et~al\mbox{.}(2024)]%
        {bao2024decoding}
\bibfield{author}{\bibinfo{person}{Keqin Bao}, \bibinfo{person}{Jizhi Zhang}, \bibinfo{person}{Yang Zhang}, \bibinfo{person}{Xinyue Huo}, \bibinfo{person}{Chong Chen}, {and} \bibinfo{person}{Fuli Feng}.} \bibinfo{year}{2024}\natexlab{}.
\newblock \showarticletitle{Decoding Matters: Addressing Amplification Bias and Homogeneity Issue for LLM-based Recommendation}.
\newblock \bibinfo{journal}{\emph{EMNLP}} (\bibinfo{year}{2024}).
\newblock


\bibitem[Bao et~al\mbox{.}(2023)]%
        {bao2023tallrec}
\bibfield{author}{\bibinfo{person}{Keqin Bao}, \bibinfo{person}{Jizhi Zhang}, \bibinfo{person}{Yang Zhang}, \bibinfo{person}{Wenjie Wang}, \bibinfo{person}{Fuli Feng}, {and} \bibinfo{person}{Xiangnan He}.} \bibinfo{year}{2023}\natexlab{}.
\newblock \showarticletitle{Tallrec: An effective and efficient tuning framework to align large language model with recommendation}. In \bibinfo{booktitle}{\emph{Proceedings of the 17th ACM Conference on Recommender Systems}}. \bibinfo{pages}{1007--1014}.
\newblock


\bibitem[Cao et~al\mbox{.}(2007)]%
        {cao2007learning}
\bibfield{author}{\bibinfo{person}{Zhe Cao}, \bibinfo{person}{Tao Qin}, \bibinfo{person}{Tie-Yan Liu}, \bibinfo{person}{Ming-Feng Tsai}, {and} \bibinfo{person}{Hang Li}.} \bibinfo{year}{2007}\natexlab{}.
\newblock \showarticletitle{Learning to rank: from pairwise approach to listwise approach}. In \bibinfo{booktitle}{\emph{Proceedings of the 24th international conference on Machine learning}}. \bibinfo{pages}{129--136}.
\newblock


\bibitem[Chao et~al\mbox{.}(2024)]%
        {chao2024make}
\bibfield{author}{\bibinfo{person}{Wen-Shuo Chao}, \bibinfo{person}{Zhi Zheng}, \bibinfo{person}{Hengshu Zhu}, {and} \bibinfo{person}{Hao Liu}.} \bibinfo{year}{2024}\natexlab{}.
\newblock \showarticletitle{Make large language model a better ranker}.
\newblock \bibinfo{journal}{\emph{arXiv preprint arXiv:2403.19181}} (\bibinfo{year}{2024}).
\newblock


\bibitem[Chen et~al\mbox{.}(2024a)]%
        {chen2024hllm}
\bibfield{author}{\bibinfo{person}{Junyi Chen}, \bibinfo{person}{Lu Chi}, \bibinfo{person}{Bingyue Peng}, {and} \bibinfo{person}{Zehuan Yuan}.} \bibinfo{year}{2024}\natexlab{a}.
\newblock \showarticletitle{Hllm: Enhancing sequential recommendations via hierarchical large language models for item and user modeling}.
\newblock \bibinfo{journal}{\emph{arXiv preprint arXiv:2409.12740}} (\bibinfo{year}{2024}).
\newblock


\bibitem[Chen et~al\mbox{.}(2019)]%
        {chen2019top}
\bibfield{author}{\bibinfo{person}{Minmin Chen}, \bibinfo{person}{Alex Beutel}, \bibinfo{person}{Paul Covington}, \bibinfo{person}{Sagar Jain}, \bibinfo{person}{Francois Belletti}, {and} \bibinfo{person}{Ed~H Chi}.} \bibinfo{year}{2019}\natexlab{}.
\newblock \showarticletitle{Top-k off-policy correction for a REINFORCE recommender system}. In \bibinfo{booktitle}{\emph{Proceedings of the twelfth ACM international conference on web search and data mining}}. \bibinfo{pages}{456--464}.
\newblock


\bibitem[Chen et~al\mbox{.}(2024b)]%
        {chen2024softmax}
\bibfield{author}{\bibinfo{person}{Yuxin Chen}, \bibinfo{person}{Junfei Tan}, \bibinfo{person}{An Zhang}, \bibinfo{person}{Zhengyi Yang}, \bibinfo{person}{Leheng Sheng}, \bibinfo{person}{Enzhi Zhang}, \bibinfo{person}{Xiang Wang}, {and} \bibinfo{person}{Tat-Seng Chua}.} \bibinfo{year}{2024}\natexlab{b}.
\newblock \showarticletitle{On Softmax Direct Preference Optimization for Recommendation}. In \bibinfo{booktitle}{\emph{The Thirty-eighth Annual Conference on Neural Information Processing Systems}} \emph{(\bibinfo{series}{NeurIPS '24})}.
\newblock


\bibitem[Dai et~al\mbox{.}(2025)]%
        {dai2025onepiece}
\bibfield{author}{\bibinfo{person}{Sunhao Dai}, \bibinfo{person}{Jiakai Tang}, \bibinfo{person}{Jiahua Wu}, \bibinfo{person}{Kun Wang}, \bibinfo{person}{Yuxuan Zhu}, \bibinfo{person}{Bingjun Chen}, \bibinfo{person}{Bangyang Hong}, \bibinfo{person}{Yu Zhao}, \bibinfo{person}{Cong Fu}, \bibinfo{person}{Kangle Wu}, {et~al\mbox{.}}} \bibinfo{year}{2025}\natexlab{}.
\newblock \showarticletitle{Onepiece: Bringing context engineering and reasoning to industrial cascade ranking system}.
\newblock \bibinfo{journal}{\emph{arXiv preprint arXiv:2509.18091}} (\bibinfo{year}{2025}).
\newblock


\bibitem[Dong et~al\mbox{.}(2023)]%
        {dong2023raft}
\bibfield{author}{\bibinfo{person}{Hanze Dong}, \bibinfo{person}{Wei Xiong}, \bibinfo{person}{Deepanshu Goyal}, \bibinfo{person}{Yihan Zhang}, \bibinfo{person}{Winnie Chow}, \bibinfo{person}{Rui Pan}, \bibinfo{person}{Shizhe Diao}, \bibinfo{person}{Jipeng Zhang}, \bibinfo{person}{Kashun Shum}, {and} \bibinfo{person}{Tong Zhang}.} \bibinfo{year}{2023}\natexlab{}.
\newblock \showarticletitle{Raft: Reward ranked finetuning for generative foundation model alignment}.
\newblock \bibinfo{journal}{\emph{arXiv preprint arXiv:2304.06767}} (\bibinfo{year}{2023}).
\newblock


\bibitem[Gao et~al\mbox{.}(2025)]%
        {gao2025sprec}
\bibfield{author}{\bibinfo{person}{Chongming Gao}, \bibinfo{person}{Ruijun Chen}, \bibinfo{person}{Shuai Yuan}, \bibinfo{person}{Kexin Huang}, \bibinfo{person}{Yuanqing Yu}, {and} \bibinfo{person}{Xiangnan He}.} \bibinfo{year}{2025}\natexlab{}.
\newblock \showarticletitle{Sprec: Self-play to debias llm-based recommendation}. In \bibinfo{booktitle}{\emph{Proceedings of the ACM on Web Conference 2025}}. \bibinfo{pages}{5075--5084}.
\newblock


\bibitem[Geng et~al\mbox{.}(2022a)]%
        {geng2022recommendation}
\bibfield{author}{\bibinfo{person}{Shijie Geng}, \bibinfo{person}{Shuchang Liu}, \bibinfo{person}{Zuohui Fu}, \bibinfo{person}{Yingqiang Ge}, {and} \bibinfo{person}{Yongfeng Zhang}.} \bibinfo{year}{2022}\natexlab{a}.
\newblock \showarticletitle{Recommendation as language processing (rlp): A unified pretrain, personalized prompt \& predict paradigm (p5)}. In \bibinfo{booktitle}{\emph{Proceedings of the 16th ACM conference on recommender systems}}. \bibinfo{pages}{299--315}.
\newblock


\bibitem[Geng et~al\mbox{.}(2022b)]%
        {P5}
\bibfield{author}{\bibinfo{person}{Shijie Geng}, \bibinfo{person}{Shuchang Liu}, \bibinfo{person}{Zuohui Fu}, \bibinfo{person}{Yingqiang Ge}, {and} \bibinfo{person}{Yongfeng Zhang}.} \bibinfo{year}{2022}\natexlab{b}.
\newblock \showarticletitle{Recommendation as Language Processing (RLP): A Unified Pretrain, Personalized Prompt \& Predict Paradigm (P5)}. In \bibinfo{booktitle}{\emph{Proceedings of the 16th ACM Conference on Recommender Systems}} \emph{(\bibinfo{series}{RecSys '22})}. \bibinfo{pages}{299–315}.
\newblock
\showISBNx{9781450392785}


\bibitem[Grattafiori et~al\mbox{.}(2024)]%
        {grattafiori2024llama}
\bibfield{author}{\bibinfo{person}{Aaron Grattafiori}, \bibinfo{person}{Abhimanyu Dubey}, \bibinfo{person}{Abhinav Jauhri}, \bibinfo{person}{Abhinav Pandey}, \bibinfo{person}{Abhishek Kadian}, \bibinfo{person}{Ahmad Al-Dahle}, \bibinfo{person}{Aiesha Letman}, \bibinfo{person}{Akhil Mathur}, \bibinfo{person}{Alan Schelten}, \bibinfo{person}{Alex Vaughan}, {et~al\mbox{.}}} \bibinfo{year}{2024}\natexlab{}.
\newblock \showarticletitle{The llama 3 herd of models}.
\newblock \bibinfo{journal}{\emph{arXiv preprint arXiv:2407.21783}} (\bibinfo{year}{2024}).
\newblock


\bibitem[Gui et~al\mbox{.}(2024)]%
        {gui2024bonbon}
\bibfield{author}{\bibinfo{person}{Lin Gui}, \bibinfo{person}{Cristina G{\^a}rbacea}, {and} \bibinfo{person}{Victor Veitch}.} \bibinfo{year}{2024}\natexlab{}.
\newblock \showarticletitle{Bonbon alignment for large language models and the sweetness of best-of-n sampling}.
\newblock \bibinfo{journal}{\emph{Advances in Neural Information Processing Systems}}  \bibinfo{volume}{37} (\bibinfo{year}{2024}), \bibinfo{pages}{2851--2885}.
\newblock


\bibitem[He et~al\mbox{.}(2017)]%
        {he2017neural}
\bibfield{author}{\bibinfo{person}{Xiangnan He}, \bibinfo{person}{Lizi Liao}, \bibinfo{person}{Hanwang Zhang}, \bibinfo{person}{Liqiang Nie}, \bibinfo{person}{Xia Hu}, {and} \bibinfo{person}{Tat-Seng Chua}.} \bibinfo{year}{2017}\natexlab{}.
\newblock \showarticletitle{Neural collaborative filtering}. In \bibinfo{booktitle}{\emph{Proceedings of the 26th international conference on world wide web}}. \bibinfo{pages}{173--182}.
\newblock


\bibitem[Jiang et~al\mbox{.}(2024)]%
        {jiang2024item}
\bibfield{author}{\bibinfo{person}{Meng Jiang}, \bibinfo{person}{Keqin Bao}, \bibinfo{person}{Jizhi Zhang}, \bibinfo{person}{Wenjie Wang}, \bibinfo{person}{Zhengyi Yang}, \bibinfo{person}{Fuli Feng}, {and} \bibinfo{person}{Xiangnan He}.} \bibinfo{year}{2024}\natexlab{}.
\newblock \showarticletitle{Item-side Fairness of Large Language Model-based Recommendation System}. In \bibinfo{booktitle}{\emph{Proceedings of the ACM on Web Conference 2024}} \emph{(\bibinfo{series}{WWW '24})}. \bibinfo{pages}{4717--4726}.
\newblock


\bibitem[Kang and McAuley(2018)]%
        {kang2018self}
\bibfield{author}{\bibinfo{person}{Wang-Cheng Kang} {and} \bibinfo{person}{Julian McAuley}.} \bibinfo{year}{2018}\natexlab{}.
\newblock \showarticletitle{Self-attentive sequential recommendation}. In \bibinfo{booktitle}{\emph{2018 IEEE international conference on data mining (ICDM)}}. IEEE, \bibinfo{pages}{197--206}.
\newblock


\bibitem[Kong et~al\mbox{.}(2025)]%
        {kong2025minionerec}
\bibfield{author}{\bibinfo{person}{Xiaoyu Kong}, \bibinfo{person}{Leheng Sheng}, \bibinfo{person}{Junfei Tan}, \bibinfo{person}{Yuxin Chen}, \bibinfo{person}{Jiancan Wu}, \bibinfo{person}{An Zhang}, \bibinfo{person}{Xiang Wang}, {and} \bibinfo{person}{Xiangnan He}.} \bibinfo{year}{2025}\natexlab{}.
\newblock \showarticletitle{MiniOneRec: An Open-Source Framework for Scaling Generative Recommendation}.
\newblock \bibinfo{journal}{\emph{arXiv preprint arXiv:2510.24431}} (\bibinfo{year}{2025}).
\newblock


\bibitem[Liao et~al\mbox{.}(2024a)]%
        {liao2024rosepo}
\bibfield{author}{\bibinfo{person}{Jiayi Liao}, \bibinfo{person}{Xiangnan He}, \bibinfo{person}{Ruobing Xie}, \bibinfo{person}{Jiancan Wu}, \bibinfo{person}{Yancheng Yuan}, \bibinfo{person}{Xingwu Sun}, \bibinfo{person}{Zhanhui Kang}, {and} \bibinfo{person}{Xiang Wang}.} \bibinfo{year}{2024}\natexlab{a}.
\newblock \showarticletitle{RosePO: Aligning LLM-based Recommenders with Human Values}.
\newblock \bibinfo{journal}{\emph{arXiv preprint arXiv:2410.12519}} (\bibinfo{year}{2024}).
\newblock


\bibitem[Liao et~al\mbox{.}(2024b)]%
        {LLaRA}
\bibfield{author}{\bibinfo{person}{Jiayi Liao}, \bibinfo{person}{Sihang Li}, \bibinfo{person}{Zhengyi Yang}, \bibinfo{person}{Jiancan Wu}, \bibinfo{person}{Yancheng Yuan}, \bibinfo{person}{Xiang Wang}, {and} \bibinfo{person}{Xiangnan He}.} \bibinfo{year}{2024}\natexlab{b}.
\newblock \showarticletitle{LLaRA: Large Language-Recommendation Assistant}. In \bibinfo{booktitle}{\emph{Proceedings of the 47th International ACM SIGIR Conference on Research and Development in Information Retrieval}} \emph{(\bibinfo{series}{SIGIR '24})}. \bibinfo{pages}{1785–1795}.
\newblock
\showISBNx{9798400704314}


\bibitem[Lin et~al\mbox{.}(2025a)]%
        {lin2025can}
\bibfield{author}{\bibinfo{person}{Jianghao Lin}, \bibinfo{person}{Xinyi Dai}, \bibinfo{person}{Yunjia Xi}, \bibinfo{person}{Weiwen Liu}, \bibinfo{person}{Bo Chen}, \bibinfo{person}{Hao Zhang}, \bibinfo{person}{Yong Liu}, \bibinfo{person}{Chuhan Wu}, \bibinfo{person}{Xiangyang Li}, \bibinfo{person}{Chenxu Zhu}, {et~al\mbox{.}}} \bibinfo{year}{2025}\natexlab{a}.
\newblock \showarticletitle{How can recommender systems benefit from large language models: A survey}.
\newblock \bibinfo{journal}{\emph{ACM Transactions on Information Systems}} \bibinfo{volume}{43}, \bibinfo{number}{2} (\bibinfo{year}{2025}), \bibinfo{pages}{1--47}.
\newblock


\bibitem[Lin et~al\mbox{.}(2025b)]%
        {lin2025order}
\bibfield{author}{\bibinfo{person}{Xinyu Lin}, \bibinfo{person}{Haihan Shi}, \bibinfo{person}{Wenjie Wang}, \bibinfo{person}{Fuli Feng}, \bibinfo{person}{Qifan Wang}, \bibinfo{person}{See-Kiong Ng}, {and} \bibinfo{person}{Tat-Seng Chua}.} \bibinfo{year}{2025}\natexlab{b}.
\newblock \showarticletitle{Order-agnostic identifier for large language model-based generative recommendation}. In \bibinfo{booktitle}{\emph{Proceedings of the 48th international ACM SIGIR conference on research and development in information retrieval}}. \bibinfo{pages}{1923--1933}.
\newblock


\bibitem[Liu et~al\mbox{.}(2025)]%
        {liu2025onerec}
\bibfield{author}{\bibinfo{person}{Zhanyu Liu}, \bibinfo{person}{Shiyao Wang}, \bibinfo{person}{Xingmei Wang}, \bibinfo{person}{Rongzhou Zhang}, \bibinfo{person}{Jiaxin Deng}, \bibinfo{person}{Honghui Bao}, \bibinfo{person}{Jinghao Zhang}, \bibinfo{person}{Wuchao Li}, \bibinfo{person}{Pengfei Zheng}, \bibinfo{person}{Xiangyu Wu}, {et~al\mbox{.}}} \bibinfo{year}{2025}\natexlab{}.
\newblock \showarticletitle{Onerec-think: In-text reasoning for generative recommendation}.
\newblock \bibinfo{journal}{\emph{arXiv preprint arXiv:2510.11639}} (\bibinfo{year}{2025}).
\newblock


\bibitem[Rafailov et~al\mbox{.}(2024)]%
        {rafailov2024direct}
\bibfield{author}{\bibinfo{person}{Rafael Rafailov}, \bibinfo{person}{Archit Sharma}, \bibinfo{person}{Eric Mitchell}, \bibinfo{person}{Christopher~D Manning}, \bibinfo{person}{Stefano Ermon}, {and} \bibinfo{person}{Chelsea Finn}.} \bibinfo{year}{2024}\natexlab{}.
\newblock \showarticletitle{Direct preference optimization: Your language model is secretly a reward model}.
\newblock \bibinfo{journal}{\emph{Advances in Neural Information Processing Systems}}  \bibinfo{volume}{36} (\bibinfo{year}{2024}).
\newblock


\bibitem[Ren et~al\mbox{.}(2024)]%
        {ren2024representation}
\bibfield{author}{\bibinfo{person}{Xubin Ren}, \bibinfo{person}{Wei Wei}, \bibinfo{person}{Lianghao Xia}, \bibinfo{person}{Lixin Su}, \bibinfo{person}{Suqi Cheng}, \bibinfo{person}{Junfeng Wang}, \bibinfo{person}{Dawei Yin}, {and} \bibinfo{person}{Chao Huang}.} \bibinfo{year}{2024}\natexlab{}.
\newblock \showarticletitle{Representation learning with large language models for recommendation}. In \bibinfo{booktitle}{\emph{Proceedings of the ACM web conference 2024}}. \bibinfo{pages}{3464--3475}.
\newblock


\bibitem[Sessa et~al\mbox{.}(2024)]%
        {sessa2024bond}
\bibfield{author}{\bibinfo{person}{Pier~Giuseppe Sessa}, \bibinfo{person}{Robert Dadashi}, \bibinfo{person}{L{\'e}onard Hussenot}, \bibinfo{person}{Johan Ferret}, \bibinfo{person}{Nino Vieillard}, \bibinfo{person}{Alexandre Ram{\'e}}, \bibinfo{person}{Bobak Shariari}, \bibinfo{person}{Sarah Perrin}, \bibinfo{person}{Abe Friesen}, \bibinfo{person}{Geoffrey Cideron}, {et~al\mbox{.}}} \bibinfo{year}{2024}\natexlab{}.
\newblock \showarticletitle{Bond: Aligning llms with best-of-n distillation}.
\newblock \bibinfo{journal}{\emph{arXiv preprint arXiv:2407.14622}} (\bibinfo{year}{2024}).
\newblock


\bibitem[Shao et~al\mbox{.}(2024)]%
        {Shao2024DeepSeekMathPT}
\bibfield{author}{\bibinfo{person}{Zhihong Shao}, \bibinfo{person}{Peiyi Wang}, \bibinfo{person}{Qihao Zhu}, \bibinfo{person}{Runxin Xu}, \bibinfo{person}{Jun-Mei Song}, \bibinfo{person}{Mingchuan Zhang}, \bibinfo{person}{Y.~K. Li}, \bibinfo{person}{Yu Wu}, {and} \bibinfo{person}{Daya Guo}.} \bibinfo{year}{2024}\natexlab{}.
\newblock \showarticletitle{DeepSeekMath: Pushing the Limits of Mathematical Reasoning in Open Language Models}.
\newblock \bibinfo{journal}{\emph{ArXiv}}  \bibinfo{volume}{abs/2402.03300} (\bibinfo{year}{2024}).
\newblock
\urldef\tempurl%
\url{https://api.semanticscholar.org/CorpusID:267412607}
\showURL{%
\tempurl}


\bibitem[Snell et~al\mbox{.}(2024)]%
        {snell2024scaling}
\bibfield{author}{\bibinfo{person}{Charlie Snell}, \bibinfo{person}{Jaehoon Lee}, \bibinfo{person}{Kelvin Xu}, {and} \bibinfo{person}{Aviral Kumar}.} \bibinfo{year}{2024}\natexlab{}.
\newblock \showarticletitle{Scaling llm test-time compute optimally can be more effective than scaling model parameters}.
\newblock \bibinfo{journal}{\emph{arXiv preprint arXiv:2408.03314}} (\bibinfo{year}{2024}).
\newblock


\bibitem[Stiennon et~al\mbox{.}(2020)]%
        {stiennon2020learning}
\bibfield{author}{\bibinfo{person}{Nisan Stiennon}, \bibinfo{person}{Long Ouyang}, \bibinfo{person}{Jeffrey Wu}, \bibinfo{person}{Daniel Ziegler}, \bibinfo{person}{Ryan Lowe}, \bibinfo{person}{Chelsea Voss}, \bibinfo{person}{Alec Radford}, \bibinfo{person}{Dario Amodei}, {and} \bibinfo{person}{Paul~F Christiano}.} \bibinfo{year}{2020}\natexlab{}.
\newblock \showarticletitle{Learning to summarize with human feedback}.
\newblock \bibinfo{journal}{\emph{Advances in neural information processing systems}}  \bibinfo{volume}{33} (\bibinfo{year}{2020}), \bibinfo{pages}{3008--3021}.
\newblock


\bibitem[Sutskever et~al\mbox{.}(2014)]%
        {sutskever2014sequence}
\bibfield{author}{\bibinfo{person}{Ilya Sutskever}, \bibinfo{person}{Oriol Vinyals}, {and} \bibinfo{person}{Quoc~V Le}.} \bibinfo{year}{2014}\natexlab{}.
\newblock \showarticletitle{Sequence to sequence learning with neural networks}.
\newblock \bibinfo{journal}{\emph{Advances in neural information processing systems}}  \bibinfo{volume}{27} (\bibinfo{year}{2014}).
\newblock


\bibitem[Tan et~al\mbox{.}(2025)]%
        {tan2025reinforced}
\bibfield{author}{\bibinfo{person}{Junfei Tan}, \bibinfo{person}{Yuxin Chen}, \bibinfo{person}{An Zhang}, \bibinfo{person}{Junguang Jiang}, \bibinfo{person}{Bin Liu}, \bibinfo{person}{Ziru Xu}, \bibinfo{person}{Han Zhu}, \bibinfo{person}{Jian Xu}, \bibinfo{person}{Bo Zheng}, {and} \bibinfo{person}{Xiang Wang}.} \bibinfo{year}{2025}\natexlab{}.
\newblock \showarticletitle{Reinforced Preference Optimization for Recommendation}.
\newblock \bibinfo{journal}{\emph{arXiv preprint arXiv:2510.12211}} (\bibinfo{year}{2025}).
\newblock


\bibitem[Wang et~al\mbox{.}(2024)]%
        {wang2024learnable}
\bibfield{author}{\bibinfo{person}{Wenjie Wang}, \bibinfo{person}{Honghui Bao}, \bibinfo{person}{Xinyu Lin}, \bibinfo{person}{Jizhi Zhang}, \bibinfo{person}{Yongqi Li}, \bibinfo{person}{Fuli Feng}, \bibinfo{person}{See-Kiong Ng}, {and} \bibinfo{person}{Tat-Seng Chua}.} \bibinfo{year}{2024}\natexlab{}.
\newblock \showarticletitle{Learnable item tokenization for generative recommendation}. In \bibinfo{booktitle}{\emph{Proceedings of the 33rd ACM International Conference on Information and Knowledge Management}}. \bibinfo{pages}{2400--2409}.
\newblock


\bibitem[Wei et~al\mbox{.}(2022)]%
        {wei2022chain}
\bibfield{author}{\bibinfo{person}{Jason Wei}, \bibinfo{person}{Xuezhi Wang}, \bibinfo{person}{Dale Schuurmans}, \bibinfo{person}{Maarten Bosma}, \bibinfo{person}{Fei Xia}, \bibinfo{person}{Ed Chi}, \bibinfo{person}{Quoc~V Le}, \bibinfo{person}{Denny Zhou}, {et~al\mbox{.}}} \bibinfo{year}{2022}\natexlab{}.
\newblock \showarticletitle{Chain-of-thought prompting elicits reasoning in large language models}.
\newblock \bibinfo{journal}{\emph{Advances in neural information processing systems}}  \bibinfo{volume}{35} (\bibinfo{year}{2022}), \bibinfo{pages}{24824--24837}.
\newblock


\bibitem[Welleck et~al\mbox{.}(2024)]%
        {welleck2024decoding}
\bibfield{author}{\bibinfo{person}{Sean Welleck}, \bibinfo{person}{Amanda Bertsch}, \bibinfo{person}{Matthew Finlayson}, \bibinfo{person}{Hailey Schoelkopf}, \bibinfo{person}{Alex Xie}, \bibinfo{person}{Graham Neubig}, \bibinfo{person}{Ilia Kulikov}, {and} \bibinfo{person}{Zaid Harchaoui}.} \bibinfo{year}{2024}\natexlab{}.
\newblock \showarticletitle{From decoding to meta-generation: Inference-time algorithms for large language models}.
\newblock \bibinfo{journal}{\emph{arXiv preprint arXiv:2406.16838}} (\bibinfo{year}{2024}).
\newblock


\bibitem[Wu et~al\mbox{.}(2024b)]%
        {surveyLLM4rec}
\bibfield{author}{\bibinfo{person}{Likang Wu}, \bibinfo{person}{Zhi Zheng}, \bibinfo{person}{Zhaopeng Qiu}, \bibinfo{person}{Hao Wang}, \bibinfo{person}{Hongchao Gu}, \bibinfo{person}{Tingjia Shen}, \bibinfo{person}{Chuan Qin}, \bibinfo{person}{Chen Zhu}, \bibinfo{person}{Hengshu Zhu}, \bibinfo{person}{Qi Liu}, \bibinfo{person}{Hui Xiong}, {and} \bibinfo{person}{Enhong Chen}.} \bibinfo{year}{2024}\natexlab{b}.
\newblock \showarticletitle{A Survey on Large Language Models for Recommendation}.
\newblock \bibinfo{journal}{\emph{World Wide Web}} \bibinfo{volume}{27}, \bibinfo{number}{5} (\bibinfo{date}{Aug.} \bibinfo{year}{2024}), \bibinfo{numpages}{31}~pages.
\newblock
\showISSN{1386-145X}


\bibitem[Wu et~al\mbox{.}(2024a)]%
        {wu2024inference}
\bibfield{author}{\bibinfo{person}{Yangzhen Wu}, \bibinfo{person}{Zhiqing Sun}, \bibinfo{person}{Shanda Li}, \bibinfo{person}{Sean Welleck}, {and} \bibinfo{person}{Yiming Yang}.} \bibinfo{year}{2024}\natexlab{a}.
\newblock \showarticletitle{Inference scaling laws: An empirical analysis of compute-optimal inference for problem-solving with language models}.
\newblock \bibinfo{journal}{\emph{arXiv preprint arXiv:2408.00724}} (\bibinfo{year}{2024}).
\newblock


\bibitem[Yang et~al\mbox{.}(2025b)]%
        {yang2025comprehensive}
\bibfield{author}{\bibinfo{person}{Hailan Yang}, \bibinfo{person}{Zhenyu Qi}, \bibinfo{person}{Shuchang Liu}, \bibinfo{person}{Xiaoyu Yang}, \bibinfo{person}{Xiaobei Wang}, \bibinfo{person}{Xiang Li}, \bibinfo{person}{Lantao Hu}, \bibinfo{person}{Han Li}, {and} \bibinfo{person}{Kun Gai}.} \bibinfo{year}{2025}\natexlab{b}.
\newblock \showarticletitle{Comprehensive list generation for multi-generator reranking}. In \bibinfo{booktitle}{\emph{Proceedings of the 48th International ACM SIGIR Conference on Research and Development in Information Retrieval}}. \bibinfo{pages}{2298--2308}.
\newblock


\bibitem[Yang et~al\mbox{.}(2025a)]%
        {yang2025breaking}
\bibfield{author}{\bibinfo{person}{Weiqin Yang}, \bibinfo{person}{Jiawei Chen}, \bibinfo{person}{Shengjia Zhang}, \bibinfo{person}{Peng Wu}, \bibinfo{person}{Yuegang Sun}, \bibinfo{person}{Yan Feng}, \bibinfo{person}{Chun Chen}, {and} \bibinfo{person}{Can Wang}.} \bibinfo{year}{2025}\natexlab{a}.
\newblock \showarticletitle{Breaking the top-k barrier: Advancing top-k ranking metrics optimization in recommender systems}. In \bibinfo{booktitle}{\emph{Proceedings of the 31st ACM SIGKDD Conference on Knowledge Discovery and Data Mining V. 2}}. \bibinfo{pages}{3542--3552}.
\newblock


\bibitem[Zhang et~al\mbox{.}(2024)]%
        {zhang2024agentcf}
\bibfield{author}{\bibinfo{person}{Junjie Zhang}, \bibinfo{person}{Yupeng Hou}, \bibinfo{person}{Ruobing Xie}, \bibinfo{person}{Wenqi Sun}, \bibinfo{person}{Julian McAuley}, \bibinfo{person}{Wayne~Xin Zhao}, \bibinfo{person}{Leyu Lin}, {and} \bibinfo{person}{Ji-Rong Wen}.} \bibinfo{year}{2024}\natexlab{}.
\newblock \showarticletitle{Agentcf: Collaborative learning with autonomous language agents for recommender systems}. In \bibinfo{booktitle}{\emph{Proceedings of the ACM Web Conference 2024}}. \bibinfo{pages}{3679--3689}.
\newblock


\bibitem[Zhang et~al\mbox{.}(2025)]%
        {zhang2025generation}
\bibfield{author}{\bibinfo{person}{Kaike Zhang}, \bibinfo{person}{Xiaobei Wang}, \bibinfo{person}{Xiaoyu Yang}, \bibinfo{person}{Shuchang Liu}, \bibinfo{person}{Hailan Yang}, \bibinfo{person}{Xiang Li}, \bibinfo{person}{Fei Sun}, {and} \bibinfo{person}{Qi Cao}.} \bibinfo{year}{2025}\natexlab{}.
\newblock \showarticletitle{From generation to consumption: Personalized list value estimation for re-ranking}.
\newblock \bibinfo{journal}{\emph{arXiv preprint arXiv:2508.02242}} (\bibinfo{year}{2025}).
\newblock


\bibitem[Zhao et~al\mbox{.}(2017)]%
        {zhao2017deep}
\bibfield{author}{\bibinfo{person}{Xiangyu Zhao}, \bibinfo{person}{Liang Zhang}, \bibinfo{person}{Long Xia}, \bibinfo{person}{Zhuoye Ding}, \bibinfo{person}{Dawei Yin}, {and} \bibinfo{person}{Jiliang Tang}.} \bibinfo{year}{2017}\natexlab{}.
\newblock \showarticletitle{Deep reinforcement learning for list-wise recommendations}.
\newblock \bibinfo{journal}{\emph{arXiv preprint arXiv:1801.00209}} (\bibinfo{year}{2017}).
\newblock


\bibitem[Ziegler et~al\mbox{.}(2005)]%
        {Ziegler2005ImprovingRL}
\bibfield{author}{\bibinfo{person}{Cai-Nicolas Ziegler}, \bibinfo{person}{Sean~M. McNee}, \bibinfo{person}{Joseph~A. Konstan}, {and} \bibinfo{person}{Georg Lausen}.} \bibinfo{year}{2005}\natexlab{}.
\newblock \showarticletitle{Improving recommendation lists through topic diversification}. In \bibinfo{booktitle}{\emph{The Web Conference}}.
\newblock


\end{thebibliography}
